\documentclass[fleqn,usenatbib,useAMS]{mnras}

\usepackage{subfiles}
\usepackage{amsmath, graphicx, amssymb,xifthen}

\usepackage[T1]{fontenc}
\usepackage{ae,aecompl}

\usepackage{newtxtext,newtxmath}

\title[Imbalanced Turbulence]{Kinetic Simulations of Imbalanced Turbulence in a Relativistic Plasma: Net Flow and Particle Acceleration}

\author[A. M. Hankla et al.]{Amelia M. Hankla,$^{1, 2}$%
\thanks{E-mail: \href{mailto:lia.hankla@gmail.com}{lia.hankla@gmail.com}, \href{mailto:amelia.hankla@colorado.edu}{amelia.hankla@colorado.edu}}%
\thanks{National Science Foundation Graduate Research Fellow}
Vladimir Zhdankin,$^{3}$ Gregory R. Werner,$^{4}$ Dmitri A. Uzdensky$^{4}$ 
\newauthor{and Mitchell C. Begelman$^{1,5}$}
\\
$^{1}$JILA, University of Colorado and National Institute of Standards and Technology, 440 UCB, Boulder, CO 80309-0440, USA\\
$^{2}$Department of Physics, University of Colorado, 390 UCB, Boulder, CO 80309-0390, USA\\
$^{3}$Department of Astrophysical Sciences, Princeton University, 4 Ivy Lane, Princeton, NJ 08544, USA\\
$^{4}$Center for Integrated Plasma Studies, Department of Physics, University of Colorado, 390 UCB, Boulder, CO 80309-0390, USA\\
${^5}$Department of Astrophysical and Planetary Sciences, University of Colorado, 391 UCB, Boulder, CO 80309, USA}

\date{Last updated \today; in original form \today}

\pubyear{2021}

\begin{document}
\label{firstpage}
\pagerange{\pageref{firstpage}--\pageref{lastpage}}
\maketitle

\begin{abstract}
Turbulent high-energy astrophysical systems often feature asymmetric energy injection: for instance, Alfv\'{e}n waves propagating from an accretion disk into its corona. Such systems are ``imbalanced”: the energy fluxes parallel and anti-parallel to the large-scale magnetic field are unequal. In the past, numerical studies of imbalanced turbulence have focused on the magnetohydrodynamic regime. In the present study, we investigate externally-driven imbalanced turbulence in a collisionless, ultrarelativistically hot, magnetized pair plasma using three-dimensional particle-in-cell (PIC) simulations. We find that the injected electromagnetic momentum efficiently converts into plasma momentum, resulting in net motion along the background magnetic field with speeds up to a significant fraction of lightspeed. This discovery has important implications for the launching of accretion disk winds. We also find that although particle acceleration in imbalanced turbulence operates on a slower timescale than in balanced turbulence, it ultimately produces a power-law energy distribution similar to balanced turbulence. Our results have ramifications for black hole accretion disk coronae, winds, and jets.
\end{abstract}



\section{Introduction} \label{sec:intro}
High-energy astrophysical systems such as accretion disks, jets, and pulsar wind nebulae often comprise collisionless, relativistically-hot plasmas and are likely turbulent~\citep{Sparks1996,Hester2008,SS1973,Balbus1998}. Turbulence in systems with magnetization (the ratio of magnetic enthalpy to plasma enthalpy)~$\sigma\gtrsim1$ can efficiently accelerate particles, as recently demonstrated in particle-in-cell (PIC) simulations~\citep{Zhdankin2017,Zhdankin2018b,Comisso2018,Comisso2019}. Such nonthermal particle acceleration (NTPA) could explain the power laws seen in spectra of jets, pulsar wind nebulae, and stellar-mass black hole X-ray binary systems~\citep{Remillard2006,Abdo2009,Aleksic2015}.

These previous studies of turbulence in relativistic collisionless plasmas have assumed symmetric energy injection into the plasma. However, this assumption is not true in a variety of space and astrophysical systems where turbulence is preferentially stirred on one side of the system. For example, in an accretion disk-wind system, turbulence in the disk may shake the footpoint of an open large-scale magnetic field line, sending Alfv\'en waves predominately away from the disk’s midplane and into the corona~\citep{Chandran2018}. This asymmetric propagation of Alfv\'en waves could impact NTPA. In addition, if efficiently coupled to the plasma, such asymmetrically-injected electromagnetic momentum could result in bulk motion of the plasma---an outflow/wind. Understanding both NTPA and the possible formation of an outflow necessitates studying turbulence with asymmetric momentum injection---so-called ``imbalanced" turbulence.

Most studies of imbalanced turbulence have focused on the magnetohydrodynamic (MHD) regime. Canonical phenomenological models for strong, ``balanced" MHD turbulence consider ensembles of counter-propagating Alfv\'en waves with equal energy fluxes along a background magnetic field~\citep{Iroshnikov1964, Kraichnan1965,GS1995,Thompson1998,Boldyrev2006}; see~\citet{Schekochihin2020} for a recent review. Phenomenological models for imbalanced turbulence relax the assumption of equal fluxes, leading to predictions that are ripe for numerical exploration~\citep{Lithwick2007,Chandran2008,Beresnyak2008,Perez2009}. Numerical attempts to model imbalanced turbulence in the MHD regime have proven difficult due to questions about the effects of varying dissipation prescriptions, limited dynamic ranges of accessible simulation domains, and limited run times~\citep{Beresnyak2009,Beresnyak2010,PB2010b,PB2010a,Perez2012,Mason2012}. Some numerical studies have extended beyond standard MHD to the relativistic MHD regime using the force-free assumption that~$\sigma\gg1$~\citep{Cho2014,Kim2015}.

Below MHD scales, analytic models of imbalanced kinetic turbulence have recently been formulated in the nonrelativistic regime~\citep{Voitenko2016,Passot2019,Gogoberidze2020}. These models of collisionless imbalanced turbulence can be tested against measurements of the solar wind~\citep{Chen2016}. Meanwhile, numerical studies have made approximations of infinite ion-to-electron mass ratio to model scales below the proton gyroradius~\citep{Cho2016}, demonstrated the importance of finite Larmor radius effects on the turbulent energy cascade~\citep{Meyrand2020}, or employed a diffusive model to study turbulence from fluid to sub-ion scales~\citep{Miloshevich2020}. A few numerical studies have modeled the fully kinetic collisionless regime with an eye towards the solar wind~\citep{Groselj2018}. However, none to our knowledge have examined the ultrarelativistic, collisionless regime relevant to high-energy astrophysical systems. Studying this regime is important because quasi-linear models of turbulent particle acceleration for imbalanced MHD predict a decrease in the Fokker-Planck momentum diffusion coefficient for increasing imbalance, leading to less efficient NTPA and posing a potential obstacle to turbulence as an astrophysical particle accelerator in some systems~\citep{Schlickeiser1989,Chandran2000}.

In this work, we explore imbalanced, relativistic turbulence in magnetized collisionless electron-positron (pair) plasmas using 3D PIC methods~\citep{Zhdankin2018}. We study how imbalance affects self-consistent NTPA, inaccessible in fluid-based models, and how it introduces an effect entirely absent in balanced models: the transfer of net momentum to the plasma, which in realistic systems could form outflows. Though the regime we simulate is particularly applicable to black hole accretion disk coronal heating~\citep{Chandran2018} and wind-launching, our results should be generally applicable to relativistic astrophysical turbulence where the source of perturbations is localized, such as the jets originating from active galactic nuclei.

To frame our study of this numerically- and analytically-unexplored regime of turbulence, we focus on four main questions: 
\begin{enumerate}
    \item How does imbalance affect the formation of a turbulent cascade?
    \item How does imbalance affect the partition of large-scale injected energy into electromagnetic, internal, and turbulent kinetic energy?
    \item Does imbalance drive net motion of the plasma?
    \item How does imbalance affect NTPA?
\end{enumerate}
We first introduce the numerical tools and parameters used to describe imbalanced turbulence (Section~\ref{sec:methods}). We then present the results of 3D collisionless, relativistic PIC simulations with varying degrees of imbalance and ratio of system size to initial Larmor radius that address each of the above questions in order. After first demonstrating the presence of a turbulent cascade (Section~\ref{ssec:turbcascade}), we discuss how and why the energy partition changes with imbalance (Section~\ref{ssec:energypartition}). Then we examine the formation of a net flow via efficient momentum transfer to the plasma and provide an analytic framework for understanding it (Section~\ref{ssec:net}). We continue by demonstrating, for the first time, the similarity of NTPA in balanced and imbalanced turbulence (Section~\ref{ssec:ntpa}). We check the dependence of our results on simulation domain size in Appendix~\ref{app:boxsize}. We conclude with implications for high-energy astrophysical systems and remaining questions (Section~\ref{sec:conclusions}).

\section{Methods}\label{sec:methods}
In this section, we will first review physical properties of relativistic magnetized plasmas (Section~\ref{ssec:background}), and then outline the simulation suite used to study imbalanced turbulence (Section~\ref{ssec:numerics}). The last subsection discusses various diagnostics that will be used to analyze our simulations (Section~\ref{ssec:diagnostics}).

\subsection{Plasma Physical Regime} \label{ssec:background}
The plasmas considered in this paper are collisionless, ultrarelativistically hot, and magnetized. This section discusses the parameters that characterize such a plasma. Throughout this work, we will use~$\langle\cdot\rangle$ to denote the time-dependent volume- or particle ensemble-average of a quantity and~$\bar\cdot$ to denote the time-average of a quantity in the time interval~$10<tc/L<20$ unless otherwise specified; here~$L/c$ is the light-crossing time of the simulation domain with length~$L$.

The parameters that characterize an ultra-relativistic, magnetized pair plasma with~$\langle\gamma\rangle\gg1$ include: an average total (electron plus positron) particle density~$n_0$, average particle energy~$\langle\gamma\rangle m_e c^2$, and characteristic magnetic field strength~$B_{\rm rms}=\sqrt{\langle B_x^2+B_y^2+B_z^2\rangle}$. Here,~$m_e$ is the mass of the electron (or positron) and~$\langle\gamma\rangle$ is the average particle Lorentz factor ($\gamma=1/\sqrt{1-v^2/c^2}=\sqrt{1+p^2/m_e^2c^2}$ for a particle with velocity~$v$, mass~$m_e$, and momentum~$p$). The fundamental physical length scales in this plasma are: the characteristic Larmor radius~$\rho_e=\langle\gamma\rangle m_ec^2/e B_{\rm rms}$, the plasma skin depth~$d_e=\left(\langle\gamma\rangle m_ec^2/(4\pi n_0 e^2)\right)^{1/2}$, and the size of the system~$L$. For a plasma with a Maxwell-J\"uttner particle distribution~$f(\gamma)=\gamma^2\sqrt{1-1/\gamma^2}\left[\theta K_2(1/\theta)\right]^{-1}\exp\left(-\gamma/\theta\right)$, the plasma has a well-defined temperature~$T_e=\langle\gamma\rangle m_ec^2/3$ (assumed equal for electrons and positrons) and the Debye length simplifies to~$\lambda_D=d_e/\sqrt{3}$. Here~$\theta\equiv T_e/m_ec^2$ and~$K_2(x)$ is the modified Bessel function of the second kind. The three length scales~$\rho_e$,~$d_e$, and~$L$ form two dimensionless quantities: the magnetization~$\sigma=B_{\rm rms}^2/4\pi h=3(d_e/\rho_e)^2/4$ and the ratio of the largest characteristic scale of spatial variation~$L/2\pi$ (which will be the turbulence driving scale in our study as described below) to the Larmor radius~$\rho_e$. Here~$h=n_0\langle\gamma \rangle m_e c^2+\langle P \rangle \approx 4n_0\langle\gamma\rangle m_e c^2/3$ is the characteristic relativistic enthalpy density and~$\langle P \rangle\approx n_0\langle\gamma\rangle m_ec^2/3$ is the (assumed isotropic) average plasma pressure. The magnetization is related to the plasma beta parameter~$\beta=8\pi \langle P\rangle/B_{\rm rms}^2$ as~$\beta=1/(2\sigma)$ and determines the relativistic Alfv\'en speed~$v_A=c\sqrt{\sigma/(\sigma+1)}$. Since we consider primarily Alfv\'enic turbulence, the magnetization governs how relativistic the large-scale turbulent motion is.

\subsection{Numerical Simulations}\label{ssec:numerics}
To explore the properties of imbalanced turbulence in a collisionless relativistic pair plasma from first principles, we use the electromagnetic PIC code {\sc Zeltron}~\citep{Cerutti2013}. {\sc Zeltron} samples the particle phase space with macro-particles and evolves them according to the Lorentz force law, providing an approximate solution to the relativistic Vlasov equation. The electric and magnetic fields evolve according to Maxwell's equations, with the addition of an externally-driven volumetric current to Amp\`ere's law to generate turbulence, as discussed below.

The physical parameters of the simulations we present are identical to those described in~\citet{Zhdankin2018} and~\citet{Zhdankin2018b} except for the modifications outlined below to introduce imbalance. Each simulation is initialized with an electron-positron plasma at rest with a Maxwell-J\"uttner distribution function, a uniform background magnetic field~${\boldsymbol B}_0=B_0\boldsymbol{\hat z}$ and no initial electromagnetic fluctuations. For each of the simulations, we set the initial magnetization to~$\sigma_0=0.5$, yielding a relativistic Alfv\'en velocity~$v_{A0}=0.58c$ and plasma beta~$\beta_0=1.0$. The initial temperature of the plasma is fixed at~$T_e=100m_ec^2$ across all simulations, corresponding to an initial average particle Lorentz factor of~$\langle\gamma\rangle\approx300$. 

To obtain the largest possible inertial range, the simulation suite's chosen numerical parameters maximize the separation between the large driving scale~$L/2\pi$ and the small initial kinetic scales~$\rho_{e0}$ and~$d_{e0}$, while still resolving the latter. We resolve the initial plasma length scales with fixed~$\Delta x=\rho_{e0}/1.5=d_{e0}/1.22$, where~$\Delta x=\Delta y=\Delta z$ is the grid cell length in each direction. The simulation domain is cubic with periodic boundary conditions and length~$L\equiv N\Delta x$, where~$N$ is the number of cells in a spatial dimension (throughout,~$N_x=N_y=N_z\equiv N$). The timestep is a fraction of the cell light-crossing time, i.e.~$\Delta t=3^{-1/2} \Delta x/c$. The simulations are initialized with 32 particles per cell per species. To scan the ratio~$L/2\pi\rho_{e0}$, we vary the number of cells in each spatial dimension, with~$N=256$, 384, 512, and 768 corresponding to~$L/2\pi\rho_{e0}\in\{27.1, 40.7, 54.3, 81.5\}$. When examining the dependence of results on simulation size, we also include three simulations of balanced turbulence with~$L/2\pi\rho_{e0}\in\{81.5, 108.7, 164\}$ ($N\in\{768, 1024, 1536\}$) used in~\citet{Zhdankin2018b} that are otherwise identical to the simulations presented in this work. 

The initial equilibrium is disrupted by an externally-driven current. We employ an oscillating Langevin antenna~\citep[OLA]{TenBarge2014} to drive turbulence volumetrically and continually throughout each simulation's duration. The OLA is implemented by adding the external current to the evolution equation for the electric field (Amp\`ere's law). This current generates counter-propagating Alfv\'en waves. The amplitudes of these counter-propagating waves are modified to induce imbalanced turbulence, as described in the following paragraphs. Because of the random nature of the OLA driving, a single simulation may not be representative of the entire ensemble of possible random seeds. To avoid basing all our conclusions on a single data point for each balance parameter, we also present a statistical study of random seeds. For each balance parameter, eight values of the random seed are simulated for the domain size~$L/2\pi\rho_{e0}=40.7$ ($N=384$). The results of the statistical study are compared against the largest simulation domains~$L/2\pi\rho_{e0}=81.5$ ($N=768$). Statistical variation could potentially be reduced in a single simulation by introducing more than eight driving modes. 

We drive imbalanced turbulence via eight independently-evolved, externally-driven sinusoidal current modes. These current modes create magnetic field perturbations propagating in opposite directions along the background magnetic field, i.e. Alfv\'en waves. The driven current modes have the form:
\begin{align}
    J_{x}^{\rm ext}({\boldsymbol x}, t)&=\frac{2\pi c}{L^2}{\rm Re}\left[\sum_{j=1}^2 \left(a_j(t) e^{i{\boldsymbol k_j\cdot \boldsymbol{x}}} + b_j(t) e^{-i{\boldsymbol k_j\cdot \boldsymbol{x}}} \right)\right]\\
    J_{y}^{\rm ext}({\boldsymbol x}, t)&=\frac{2\pi c}{L^2}{\rm Re}\left[\sum_{j=3}^4 \left(a_j(t) e^{i{\boldsymbol k_j\cdot \boldsymbol{x}}} + b_j(t) e^{-i{\boldsymbol k_j\cdot \boldsymbol{x}}} \right)\right]\\
    J_{z}^{\rm ext}({\boldsymbol x}, t)&=\frac{2\pi c}{L^2}{\rm Re}\left[\sum_{j=1}^4 \left(-a_j(t) e^{i{\boldsymbol k_j\cdot \boldsymbol{x}}} + b_j(t) e^{-i{\boldsymbol k_j\cdot \boldsymbol{x}}} \right) \right]. 
\end{align}
The sign of~$k_z$ dictates the direction of the current mode's propagation. Four of the modes have no~$y$-component of their wavevector and four have no~$x$-component; four propagate in the~$+z$-direction and four propagate in the~$-z$-direction. These wavevectors are
\begin{align}
    \boldsymbol{k_1}&=k_0(-1,0,1) & \boldsymbol{k_2}&=k_0(1,0,1)\\
    \boldsymbol{k_3}&=k_0(0,-1,1) & \boldsymbol{k_4}&=k_0(0,1,1).
\end{align}
Here,~$k_0=2\pi/L$, so that the driving scale is the largest scale~$L/2\pi$. We ensure~$\nabla \cdot {\boldsymbol J}_{\rm ext}=0$ to avoid local injection of net charge. Currents driven in~$J_{{\rm ext}, x}$ and~$J_{{\rm ext}, y}$ create Alfv\'en waves with magnetic field perturbations in the~$y$- and~$x-$directions, respectively. The amplitudes of these currents can be adjusted to create counter-propagating Alfv\'en waves of unequal amplitudes, thus enabling our study of imbalanced turbulence.

The external current's time-dependence is dictated by the coefficients~$a_j(t)$ and~$b_j(t)$. The coefficient at the ($n$+1)th timestep is found from the previous~$n$th timestep as 
\begin{align}
    a_j^{(n+1)}=a_j^{(n)} e^{-i\omega \Delta t}+\alpha_j u_j^{(n)}\Delta t\label{eq:aj}\\
    b_j^{(n+1)}=b_j^{(n)} e^{-i\omega \Delta t}+\beta_j v_j^{(n)}\Delta t.\label{eq:bj}
\end{align}
The coefficients~$a_j(t)$ and~$b_j(t)$ thus oscillate at frequency~$\omega$ with random kicks at each timestep~\citep[cf. Langevin equation, hence the name ``oscillating Langevin antenna'';][]{TenBarge2014}. The initial coefficients~$a_j^{(0)}$ and~$b_j^{(0)}$ are set to amplitudes~$\mathcal{A}$ and~$\mathcal{B}$ multiplied by random phases~$\phi_j^{(a)}$ and~$\phi_j^{(b)}$:~$a_j^{(0)}=\mathcal{A}e^{i\phi_j^{(a)}}$ and~$b_j^{(0)}=\mathcal{B}e^{i\phi_j^{(b)}}$. We set~$\mathcal{A}=B_0L/8\pi$, which for balanced turbulence achieves~$\delta B_{\rm rms}=\sqrt{B_{\rm rms}^2-B_0^2}\sim B_0$. The random kicks~$u_j^{(n)}$ and~$v_j^{(n)}$ in Equations~\ref{eq:aj} and~\ref{eq:bj} are complex random numbers with real and imaginary components drawn from a uniform distribution between~$-0.5$ and 0.5. The constant parameters~$\alpha_j$ and~$\beta_j$ are set such that when ensemble-averaged,~$\langle |a_j^{(n)}|^2\rangle=\mathcal{A}^2$ and~$\langle |b_j^{(n)}|^2\rangle=\mathcal{B}^2$. The complex driving frequency~$\omega$ has real component~$\omega_0$ and an imaginary component~$-\Gamma_0$ which we set to be non-integer multiples of the Alfv\'en frequency~$\omega_A\equiv2\pi v_A/L$ to avoid initial resonances:~$\omega_0=(0.6/\sqrt{3})\omega_A\approx0.35\omega_A$ and~$\Gamma_0=(0.5/\sqrt{3})\omega_A\approx 0.29\omega_A$. In frequency space, the driving is a Lorentzian centered at~$\omega_0$ with a full-width half-max of~$\Gamma_0$; see~\citet{TenBarge2014} for details. The amplitudes~$\mathcal{A}$ and~$\mathcal{B}$ (and therefore the~$\alpha_j$ and~$\beta_j$ values) are the same for all~$a_j$ and~$b_j$, but the random parameters~$u_j$,~$v_j$, and the initial phases~$\phi_j^{(a,b)}$ are different for each~$k_j$.

We introduce imbalance by adjusting the amplitudes of the currents propagating in the~$-z$-direction relative to those propagating in the~$+z$-direction. The coefficients~$a_j$ control the currents propagating in the~$+z$-direction, whereas~$b_j$ control the waves propagating in the~$-z$-direction. These currents' amplitudes are dictated by their respective~$\mathcal{A}$ and~$\mathcal{B}$ amplitudes. To achieve imbalanced turbulence, we hold~$\mathcal{A}$ fixed and vary~$\mathcal{B}$. We quantify how balanced the turbulence is via the {\it balance parameter},
\begin{equation}
    \xi\equiv \mathcal{B}/\mathcal{A},\label{eq:balanceParam}
\end{equation}
where~$\mathcal{B}$ is the amplitude for the~$-z$-modes and~$\mathcal{A}$ is the fixed amplitude for the~$+z$-modes. A value of~$\xi=1$ corresponds to the canonical balanced case, whereas~$\xi=0$ corresponds to current modes propagating only in the~$+z$-direction. Because we drive currents rather than Alfv\'en modes, we do not directly control the exact amplitude of counter-propagating Alfv\'en waves. If~$\xi=0$ corresponded exactly to the case of Alfv\'en waves propagating in a single direction, we would not expect turbulence to develop in a non-relativistic, ideal MHD plasma. Indeed, turning off the random kicks by setting~$\alpha_j=\beta_j=0$ does not result in turbulence for simulations with~$\xi=0$ (not shown). However, with our set-up of nonzero~$\alpha_j$ and~$\beta_j$, the~$\xi=0$ case does become turbulent because the OLA forcing excites counter-propagating Alfv\'en waves. In principle, more imbalanced turbulence should be achievable by, e.g., a decaying turbulence problem or by changing the driving mechanism. For the present work, we simply term the~$\xi=0$ case the ``most imbalanced" case.

We will use the balance parameter~$\xi$ throughout this paper to refer to the degree of imbalance; however, since~$\xi$ measures the imbalance of the driving mechanism rather than the turbulence, we now briefly discuss the relationship of~$\xi$ to other methods of measuring imbalance. Cross-helicity, an invariant in ideal MHD, measures the difference in the energy densities associated with waves propagating anti-parallel (with energy density~$\mathcal{E}_+$ and amplitude~$\delta B_+$) and parallel to the magnetic field (with energy density~$\mathcal{E}_-$ and amplitude~$\delta B_-$). In ideal, non-relativistic, incompressible MHD,~$\mathcal{E}_\pm=\langle\rho |{\boldsymbol z_\pm}|^2\rangle/4$, where~${\boldsymbol z_\pm}=\boldsymbol{\delta v}\pm{\boldsymbol b}$ are the Elsasser fields~\citep{Elsasser1950}. Here~$\boldsymbol{\delta v}$ is the fluctuating plasma velocity,~${\boldsymbol b} \equiv \boldsymbol{\delta B}~v_A/B_0$ is the fluctuating magnetic field in velocity units, and~$\rho$ is the plasma mass density (not to be confused with the Larmor radius~$\rho_e$). The total energy density is then given by~$\mathcal{E}\equiv \mathcal{E}_++\mathcal{E}_-=(1/2)\langle\rho\left(|{\boldsymbol{\delta v}}|^2+|{\boldsymbol b}|^2\right)\rangle$ and the cross-helicity~$H_c\equiv \left(\mathcal{E}_+-\mathcal{E}_-\right)/\langle\rho\rangle$ can be re-expressed as~$H_c=\langle{\boldsymbol{\delta v}}\cdot{\boldsymbol b}\rangle$. Cross-helicity is related to the volume-averaged~$z$-component of the Poynting flux~${\boldsymbol S}({\boldsymbol x}, t)=(c/4\pi)\left[{\boldsymbol E}\times{\boldsymbol B}\right]$ under the assumptions of incompressible, nonrelativistic, ideal reduced MHD that~$\delta B\ll B_0$ and that the fluctuations~${\boldsymbol \delta}{\boldsymbol B}$ and~${\boldsymbol{ \delta v}}$ are perpendicular to the background field~$B_0\boldsymbol{\hat z}$:
\begin{equation}
\langle S_z\rangle(t)=-\frac{B_0}{4\pi}\langle{\boldsymbol{\delta v}}\cdot\boldsymbol{\delta B}\rangle=-\frac{1}{4\pi}\frac{B_0^2}{v_A} H_c(t).\label{eq:szHc}
\end{equation}
For a single Alfv\'en wave,~${\boldsymbol{ \delta v}}/v_A=\pm{\boldsymbol \delta}{\boldsymbol B}/B_0$ and thus the magnitude of the Poynting flux for a single Alfv\'en wave~$|\langle S_z\rangle_{1-\rm wave}|$ is
\begin{equation}
    |\langle S_z\rangle_{1-\rm wave}|=\frac{1}{4\pi}\langle\delta B^2\rangle v_A.\label{eq:maxSz}
\end{equation}
We can estimate the values of the driven Alfv\'en wave energies in our simulations of imbalanced turbulence as~$E_+\sim \langle\delta B_+^2\rangle\sim |a_j|^2$ (where~$\delta B_+$ is the amplitude of the magnetic perturbation travelling in the~$+z$ direction) and~$E_-\sim |b_j|^2$, leading to 
\begin{align}
    \langle S_z\rangle&\propto H_c\propto 1-\xi^2. \label{eq:szhcscaling}
\end{align}
Equation~\ref{eq:szhcscaling} will be tested in Section~\ref{sssec:netnumerics}. Normalizing to the total energy, the ``normalized cross-helicity"~$\tilde H_c=(\mathcal{E}_+-\mathcal{E}_-)/(\mathcal{E}_++\mathcal{E}_-)$~\citep{Perez2009, Chen2016, Meyrand2020} can be estimated as~$\tilde H_c\sim (1-\xi^2)/(1+\xi^2)$. Finally, we calculate the ratio~$\langle \overline{z_-^2}\rangle/\langle \overline{z_+^2}\rangle$ of Elsasser field energies for several~$\xi$ (Table~\ref{tab:xiZenergies}). These fields are calculated with~$v_A(t)=c\sqrt{\sigma(t)/(\sigma(t)+1)}$ using the instantaneous magnetization. For our most imbalanced turbulence ($\xi=0.0$), the ratio of energies is about~$0.72$, whereas for perfectly imbalanced turbulence the ratio would be zero. The discrepancy between the driving's imbalance and the turbulence's imbalance is due to the excitation of counter-propagating Alfv\'en waves (discussed in the previous paragraph), and possibly relativistic, kinetic, and moving frame effects, which the Elsasser fields we use do not take into account.

The main goal of our study is to determine the impact of imbalance on the properties of collisionless turbulence. We do so by varying the balance parameter~$\xi$ between 0 (``most imbalanced") and 1 (``balanced") at every value of~$L/2\pi\rho_{e0}$.
\begin{table}
\centering
\caption{Measured values of the Elsasser fields' energy ratio $r_E=\langle \overline{z_-^2}\rangle/\langle \overline{z_+^2}\rangle$ for a sampling of balance parameter~$\xi$ values. The first number gives the ratio for $N=768$ and the second gives one standard deviation of the $N=384$ statistical seed studies. Time averages are taken over~$5.0<tc/L<20.0$.}\label{tab:xiZenergies}
\begin{tabular}{cccccc}
$\xi$ & $0.0$ & $0.5$ & $1.0$ \\ \hline\hline
$r_E$ & $0.72\pm 0.06$ & $0.92\pm 0.14$ & $1.01\pm0.14$ \\
\end{tabular}
\end{table}%

\subsection{Energy Diagnostics} \label{ssec:diagnostics}
In this section, we discuss diagnostics that will be used in Section~\ref{sec:results} to partition the energy of the system into four main types. 

The total energy density in the system can be decomposed into the energy density~$\mathcal{E}_{\rm EM}$ in the electric and magnetic fields and the total (kinetic plus rest mass) energy density~$\mathcal{E}_{\rm pl}$ in the plasma particles. Fluid quantities provide intuition into the plasma's behavior by partitioning~$\mathcal{E}_{\rm pl}$ into internal, net flow, and turbulent flows:
\begin{equation}
    \mathcal{E}_{\rm pl}({\boldsymbol x},t)=\mathcal{E}_{\rm int}({\boldsymbol x},t)+\mathcal{E}_{\rm net}(t)+\mathcal{E}_{\rm turb}({\boldsymbol x},t).
\end{equation}
The plasma's total kinetic, internal, and turbulent kinetic energy densities are calculated for each simulation cell from both the electron and positron macro-particles' positions and momenta, though they will often be discussed in terms of their volume averages~$\langle\mathcal{E}_{\rm pl}\rangle$,~$\langle\mathcal{E}_{\rm int}\rangle$ and~$\langle\mathcal{E}_{\rm turb}\rangle$, respectively. The net flow energy density~$\mathcal{E}_{\rm net}$ is a key quantity for characterizing how efficiently imbalanced turbulence can drive a directed plasma flow. It is a global quantity calculated from the total momentum in the system. Explicitly, these quantities are defined as:
\begin{align}
    \mathcal{E}_{\rm int}({\boldsymbol x},t)&\equiv\sqrt{\mathcal{E}_{\rm pl}^2({\boldsymbol x},t)-\boldsymbol{\mathcal{P}}_{\rm pl}^2({\boldsymbol x},t) c^2}\label{eq:eint}\\
    \mathcal{E}_{\rm net}(t)&\equiv\langle\mathcal{E}_{\rm pl}\rangle(t) - \sqrt{\langle\mathcal{E}_{\rm pl}\rangle^2(t)-\langle\boldsymbol{\mathcal{P}}_{\rm pl}\rangle^2(t) c^2}\label{eq:enet}\\
    \mathcal{E}_{\rm turb}({\boldsymbol x},t)&\equiv\mathcal{E}_{\rm flow}({\boldsymbol x},t)-\mathcal{E}_{\rm net}(t)\label{eq:eturb}
\end{align}
where~$\mathcal{E}_{\rm flow}({\boldsymbol x},t)=\mathcal{E}_{\rm pl}({\boldsymbol x},t)-\mathcal{E}_{\rm int}({\boldsymbol x},t)$~\citep[Eq. 8]{Zhdankin2018} and~$\boldsymbol{\mathcal{P}}_{\rm pl}({\boldsymbol x},t)$ is the local momentum density of the electron-positron plasma. This framework is similar to that in~\citet{Zhdankin2018}, with the renaming of the ``bulk" energy density to the ``flow" kinetic energy density and further breakdown of the flow energy density into the energy density associated with the net flow of the plasma~$\mathcal{E}_{\rm net}$ through the simulation domain and the turbulent motions~$\mathcal{E}_{\rm turb}$. Equation~\ref{eq:eint} for the internal energy density is analogous to the relativistic energy~$E$ of a single particle~$E^2=(mc^2)^2+p^2c^2$. In this analogy, the plasma internal energy acts like a particle rest mass, the plasma momentum acts as a particle momentum, and the total plasma kinetic energy acts like a relativistic particle mass. Equation~\ref{eq:enet} for the net flow energy density uses a similar analogy, specifically applied to the quantity volume-averages.

The change in the energy of the plasma and the electromagnetic fields comes from the energy injected into the system by the OLA driving. The energy injection rate~$\langle\dot{\mathcal{E}}_{\rm inj}\rangle=-\langle{\boldsymbol J}_{\rm ext}\cdot{\boldsymbol E}\rangle$ is statistically constant in time. Integrating it over time gives the total injected energy density up to time~$t$:~$\mathcal{E}_{\rm inj}(t)$. Because the injected energy depends on the amplitude of the driven waves, its value at any given time varies with~$\xi$ within a factor of two or so (see Section~\ref{ssec:energypartition}). To account for this dependence of injected energy on balance parameter, we normalize the volume-averaged changes in the various energy components by the volume-averaged injected energy to obtain energy efficiencies: 
\begin{equation}
    1=\frac{\Delta\langle\mathcal{E}_{\rm int}\rangle}{\langle\mathcal{E}_{\rm inj}\rangle} + \frac{\Delta\mathcal{E}_{\rm net}}{\langle\mathcal{E}_{\rm inj}\rangle} + \frac{\Delta\langle\mathcal{E}_{\rm turb}\rangle}{\langle\mathcal{E}_{\rm inj}\rangle}+\frac{\Delta\langle\mathcal{E}_{\rm EM}\rangle}{\langle\mathcal{E}_{\rm inj}\rangle}, \label{eq:defEfficiency}
\end{equation}
where the terms on the right are the internal efficiency, net flow efficiency, turbulent kinetic efficiency, and electromagnetic efficiency, respectively. We use~$\Delta\mathcal{E}$ to indicate the change in a type of energy density since the start of the simulation, i.e.~$\Delta\mathcal{E}(t)=\mathcal{E}(t)-\mathcal{E}(0)$.

\section{Results} \label{sec:results}

In this section, we investigate how imbalanced turbulence differs from balanced turbulence through a series of comparisons. After demonstrating the presence of a turbulent cascade for all values of balance parameter (Section~\ref{ssec:turbcascade}), we examine how the injected energy transforms into the plasma's internal and turbulent energy (Section~\ref{ssec:energypartition}). We next turn to the novel aspect of imbalanced turbulence: the presence of a net flow (Section~\ref{ssec:net}). By using the statistical study of eight random seeds at smaller simulation domains~$L/2\pi\rho_{e0}=40.7$ to enhance the trends in the largest simulation domains~$L/2\pi\rho_{e0}=81.5$, we constrain the dependence of each of these energy types on balance parameter. We then explore the decomposition of the plasma energy into thermal and nonthermal components and how particle acceleration depends on the balance parameter (Section~\ref{ssec:ntpa}). The influence of simulation domain size on the fluid quantities is explored by varying~$L/2\pi\rho_{e0}$ in Appendix~\ref{app:boxsize}.%

\subsection{Formation of a turbulent cascade} \label{ssec:turbcascade}
\begin{figure*}
    \centering
    \includegraphics[width=.9\textwidth]{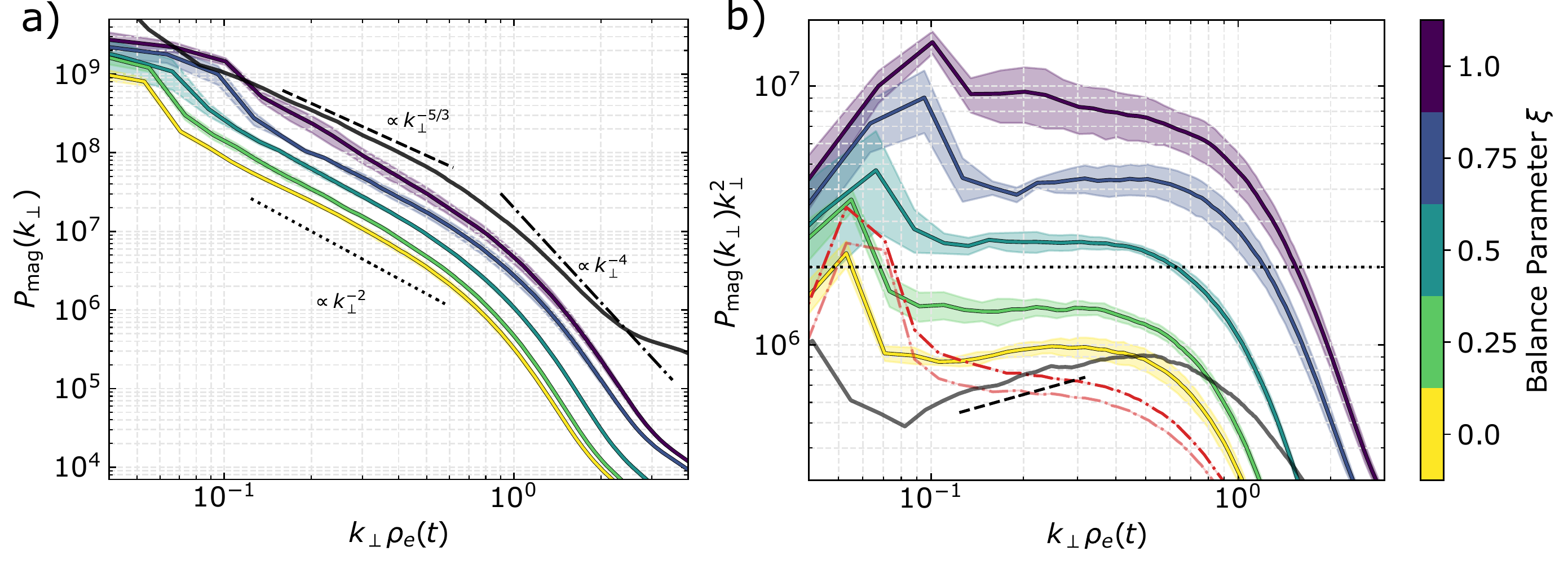}
    \caption{A turbulent cascade forms for all balance parameters. a) The magnetic energy spectra~$P_{\rm mag}(k_\perp)$ for~$L/2\pi\rho_{e0}=81.5$ simulations of varying balance parameter averaged between times~$8.8<tc/L<9.9$ (comprising five outputs) show an inertial range between~$k_\perp\rho_e(t)\sim0.1$ and~$1.0$. A break in the spectrum at~$k_\perp\rho_e(t)\sim1.0$ indicates the onset of kinetic effects. b) When compensated by~$k_\perp^{2}$, the spectra for the balanced~$\xi=0.75$ and~$1.0$ cases are slightly steeper than~$\propto k_\perp^{-2}$, whereas the imbalanced case~$\xi=0.0$ is slightly flatter. The Elsasser fields' spectra, shown in dash-dot red lines for $\xi=0$, exhibit slightly different slopes, with the stronger field ($z_+$, top line) being slightly steeper than the weaker field ($z_-$, bottom line). In both panels, shaded lines show one temporal standard deviation about the mean. Black dashed lines show the scaling~$k_\perp^{-5/3}$; black dotted lines show~$k_\perp^{-2}$. Gray lines show the~$L/2\pi\rho_{e0}=164$ balanced simulation's magnetic energy spectrum, taken at~$t=8.9~L/c$.}
    \label{fig:me-spectrum}
\end{figure*}%
The spectrum of the turbulent magnetic energy is a common diagnostic when examining turbulence. Much of the previous work on imbalanced turbulence in MHD plasmas has examined the power-law indices of the two Elsasser fields and how they may or may not deviate from Goldreich-Sridhar~$k_\perp^{-5/3}$ scalings~\citep{GS1995, Lithwick2007, Beresnyak2008}. In this study, we simply calculate the overall turbulent magnetic energy spectrum via:
\begin{equation}
    P_{\rm mag}(k_\perp,t)=\int~{\rm d}k_z{\rm d}\phi~k_\perp~\frac1{8\pi}\left(|\tilde B_x|^2+|\tilde B_y|^2+|\Tilde{\delta B_z}|^2\right),\label{eq:turbspectrum}
\end{equation}
where~$\delta B_z=B_z-B_0$,~$k_z$ are the parallel wavenumbers,~$\phi$ are the azimuthal angles, and~$\tilde\cdot$ indicates the Fourier transform.

We find that the magnetic energy spectrum shows the formation of a turbulent cascade for all balance parameters (Fig.~\ref{fig:me-spectrum}a). The spectra averaged over the time interval~$8.8<tc/L<9.9$ (corresponding to~$5.1<tv_{A0}/L<5.7$), i.e. after the turbulent cascade has fully developed but before the plasma's heating has diminished the inertial range, show similar shapes for all values of the balance parameter. The inertial range forms between~$k_\perp\rho_e(t)\sim0.08$ and~$0.6$ for the most imbalanced case; a slightly shifted inertial range beginning at~$k_\perp\rho_e(t)\sim 0.15$ rather than~$k_\perp\rho_e(t) \sim 0.08$ for the simulations with more balanced turbulence ($\xi=1.0$ and 0.75) results from the faster heating at these balance parameters. The power-law index in the inertial range is roughly consistent with~$-5/3$, the classic MHD prediction for strong turbulence~\citep{GS1995}. Although the magnetic energy spectrum better matches the $k_\perp^{-2}$ scaling characteristic of weak turbulence in the non-relativistic~\citep{Galtier2000} and relativistic~\citep{TenBarge2021, Ripperda2021} regimes, the turbulence in our simulations is strong. The steeper than~5/3 spectrum is likely due to a small domain size, as found by~\citet{Zhdankin2018}. Identical balanced simulations with twice the domain size are consistent with a power-law index of~$-5/3$ (dark grey line in Fig.~\ref{fig:me-spectrum}a). Accurately measuring the power-law indices of imbalanced turbulence via larger domain sizes is beyond the scope of the present study. Below the characteristic Larmor radius ($k_\perp\rho_e\gtrsim1$), the spectrum steepens to another power law that covers a more limited range between~$k_\perp\rho_e(t)\sim1$ and~$2$ and is broadly consistent with the formation of a kinetic cascade with a power-law index of~$-4$ for all values of imbalance~\citep[also found in][]{Zhdankin2018}, much steeper than in the inertial range. Again, however, providing more exact values to test against the predictions and measurements in~\citet{Schekochihin2009} or~\citet{Zhdankin2018} would require larger, better-resolved simulation domains. Numerical noise dominates at scales smaller than~$k_\perp\rho_e(t)\sim2$.

Although determining the precise dependence of the inertial range's slope on balance parameter would require a larger inertial range, there are hints that the slope depends on~$\xi$. When the magnetic energy spectrum is compensated by the scaling~$k_\perp^{-2}$, the simulation with balanced turbulence ($\xi=1.0$) has a downward-sloping spectrum, whereas the simulation of turbulence with~$\xi=0$ has a slightly positive slope (Fig.~\ref{fig:me-spectrum}b). However, this steepening is most likely due to increased damping. The balanced simulations heat up faster (as discussed in Section~\ref{ssec:energypartition}), resulting in smaller values of~$L/\rho_e$ than the imbalanced simulations. Due to the small domain sizes, our simulations cannot distinguish differences in slope caused by imbalance or dissipation.

Because previous MHD predictions for imbalanced turbulence generally discuss the spectra of the Elsasser fields rather than the magnetic energy spectra, we also plot the Elsasser fields' spectra for the~$\xi=0$ case (dash-dot red lines in Fig.~\ref{fig:me-spectrum}b). The larger-amplitude field appears to have a slightly steeper slope than the smaller-amplitude field, consistent with previous MHD simulations~\citep{PB2010a, Perez2012}. However, constraining the variation in the power-law index is difficult to quantify with such a short inertial range. If the dependence of the Elsasser fields' energy spectra on imbalance persists in larger simulations, it could support MHD predictions that the spectra's power-law indices depend on imbalance~\citep{Galtier2000,Chandran2008}.

\begin{figure}
    \centering
    \includegraphics[width=0.45\textwidth]{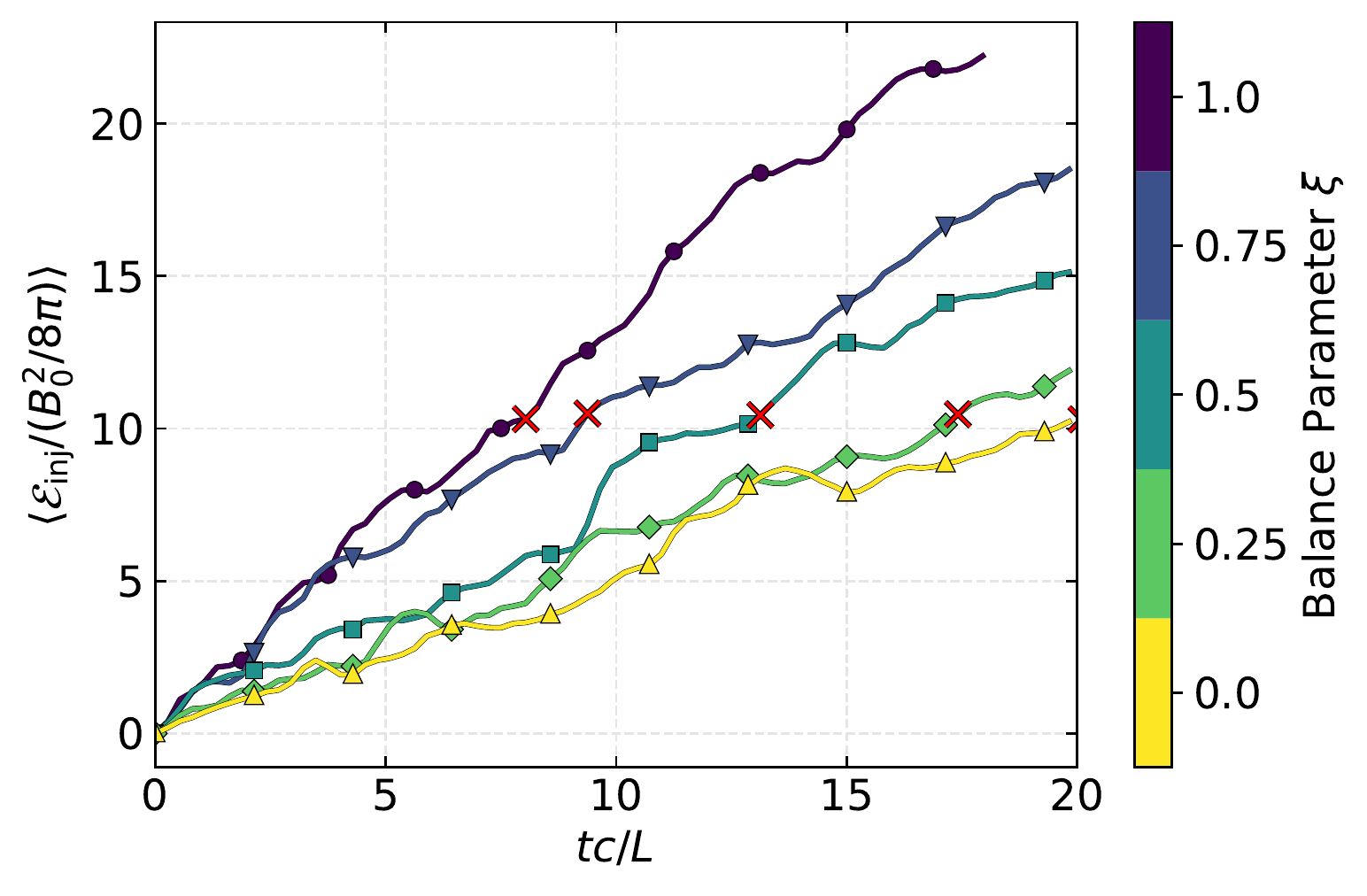}
    \caption{The amount of energy injected into a simulation depends on its balance parameter. The simulations of more balanced turbulence (purple and blue) have more injected energy than the simulations of less balanced turbulence (yellow and green). Red~$\times$'s indicate the ``equivalent" times where the same amount of energy has been injected for each simulation (see Table~\ref{tab:teq}), which all have~$L/2\pi\rho_{e0}=81.5$.}
    \label{fig:einj-evolution}
\end{figure}

\begin{figure*}
    \centering
    \includegraphics[width=0.75\textwidth]{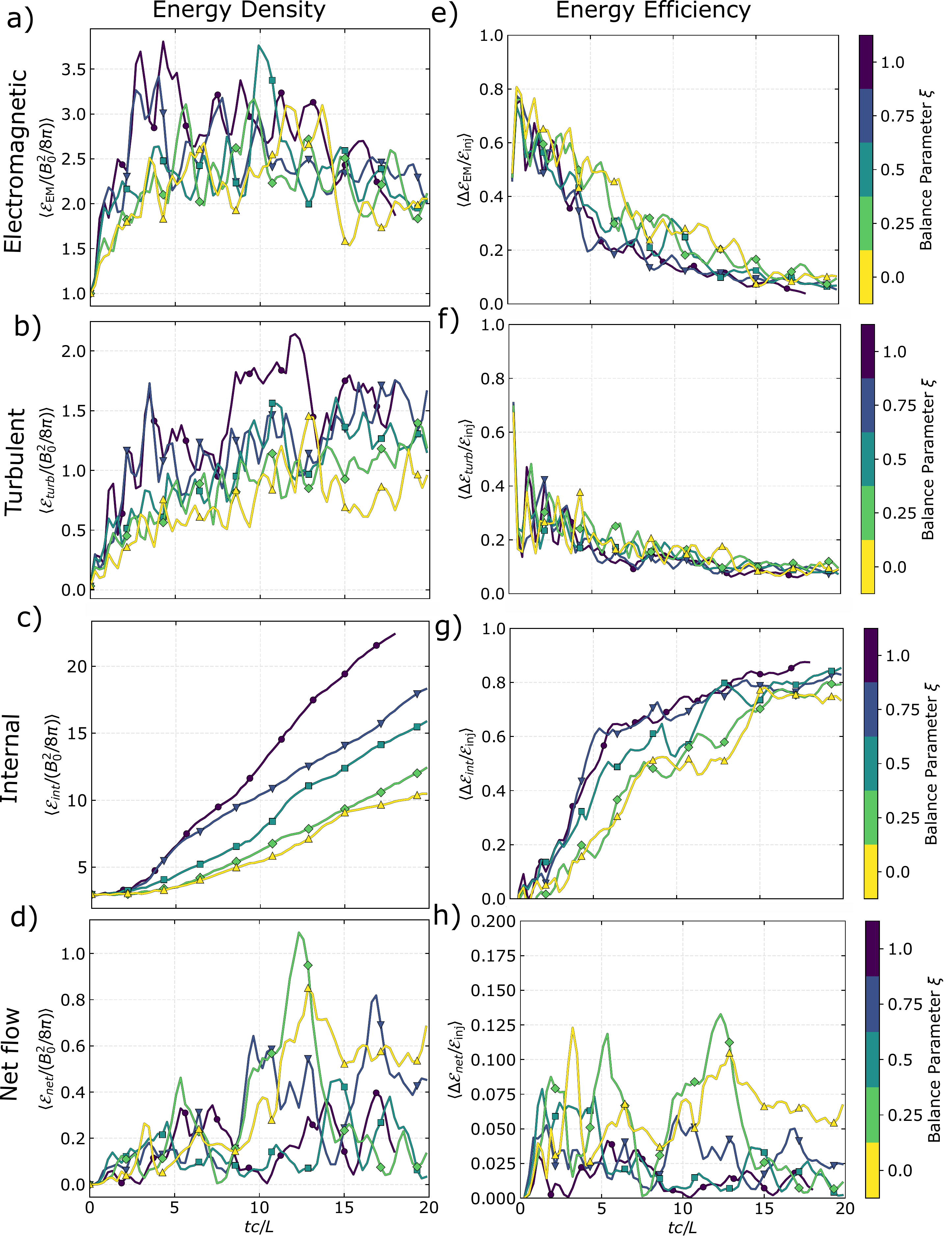}
    \caption{Energy partition into electromagnetic, turbulent kinetic, internal, and net flow energy depends on balance parameter. Left column: each type of energy density evolved over time, normalized to the constant value of the initial magnetic energy density~$B_0^2/8\pi$. The turbulent electromagnetic (a) and kinetic (b) energy densities reach a constant value whereas the internal (c) and net flow (d) energy densities increase over time. Right column: the change in each type of energy density evolved over time, normalized to the total amount of injected energy density~$\mathcal{E}_{\rm inj}(t)$. Summing over the four panels on the right for each simulation adds to~1. Turbulent electromagnetic (e) and kinetic (f) energy efficiencies decay as~$\propto t^{-1}$, whereas internal (g) and net flow (h) energy efficiencies saturate at a constant fraction of the injected energy. Note that the net flow energy (h) has a different vertical axis. Colors and markers indicate balance parameter. These simulations all have~$L/2\pi\rho_{e0}=81.5$.}
    \label{fig:grid}
\end{figure*}

Further evidence for the formation of a turbulent cascade comes from comparing the evolution of the energy injected into the simulation and the internal energy of the plasma. Whereas the accumulated injected energy increases linearly from~$t=0$ onward (Fig.~\ref{fig:einj-evolution}), the internal energy density does not begin to increase until about~$2.5~L/c$, close to one Alfv\'en-crossing time (see Section~\ref{sssec:int}, Fig.~\ref{fig:grid}c). Presumably the energy injected at the driving scale cascades to smaller scales over the time period~$t=0-2.5~L/c$ until it reaches the characteristic Larmor radius and dissipates into internal energy---i.e. the turbulent cascade forms in the first couple of light-crossing times. There appears to be an increase in the cascade formation time for decreasing balance parameter (see Section~\ref{sssec:int}). 

\subsection{Partition of the Injected Energy} \label{ssec:energypartition}
\subsubsection{Framework for the Energy Partition}\label{sssec:framework}
Because we drive the plasma in each simulation without an energy sink, the overall energy of each case increases in time. By adding a statistically-constant amount of energy at each timestep, the overall amount of injected energy increases linearly in time (Fig.~\ref{fig:einj-evolution}). The amplitude of the driven waves by definition depends on the balance parameter~$\xi$ (Equation~\ref{eq:balanceParam}), and so the amount of energy injected also depends on~$\xi$. The increase in injected energy is twice as fast for the balanced case~$\xi=1.0$ as for the most imbalanced case~$\xi=0.0$ (Fig.~\ref{fig:einj-evolution}). This doubling of injected energy occurs because twice as many modes are driven in the balanced as in the most imbalanced case. 

The injected energy converts into various types of plasma energy, each of which will be discussed in the following subsections. Fig.~\ref{fig:grid} shows temporal evolution of each quantity's energy density (left) and energy efficiency (right). The large-scale injected energy cascades to smaller scales in the form of bulk kinetic and electromagnetic energy until it is dissipated into internal energy, thereby implying that the internal energy should increase linearly in time during the statistical steady state---and it does, as shown in Fig.~\ref{fig:grid}c. As stationary conduits for the turbulent Alfv\'enic cascade, we expect both the turbulent kinetic and magnetic perturbations to fluctuate around steady-state values rather than continually increasing over time. Simulations support this idea of statistically steady-state values: the turbulent electromagnetic and kinetic energies saturate to their mean values around~$7.5~L/c$ and fluctuate thereafter around these values (Fig.~\ref{fig:grid}a, b). These values' dependence on balance parameter will be explored in Section~\ref{sssec:constants}, while the dependence of the internal energy density's slope on balance parameter is discussed in Section~\ref{sssec:int}. The final component of plasma energy, the kinetic energy of net motion through the simulation domain, does not have an easily characterized evolution (Fig.~\ref{fig:grid}d); its dependence on balance parameter will be discussed in Section~\ref{sssec:netnumerics}.

\begin{figure}
    \centering
    \includegraphics[width=0.5\textwidth,clip=true]{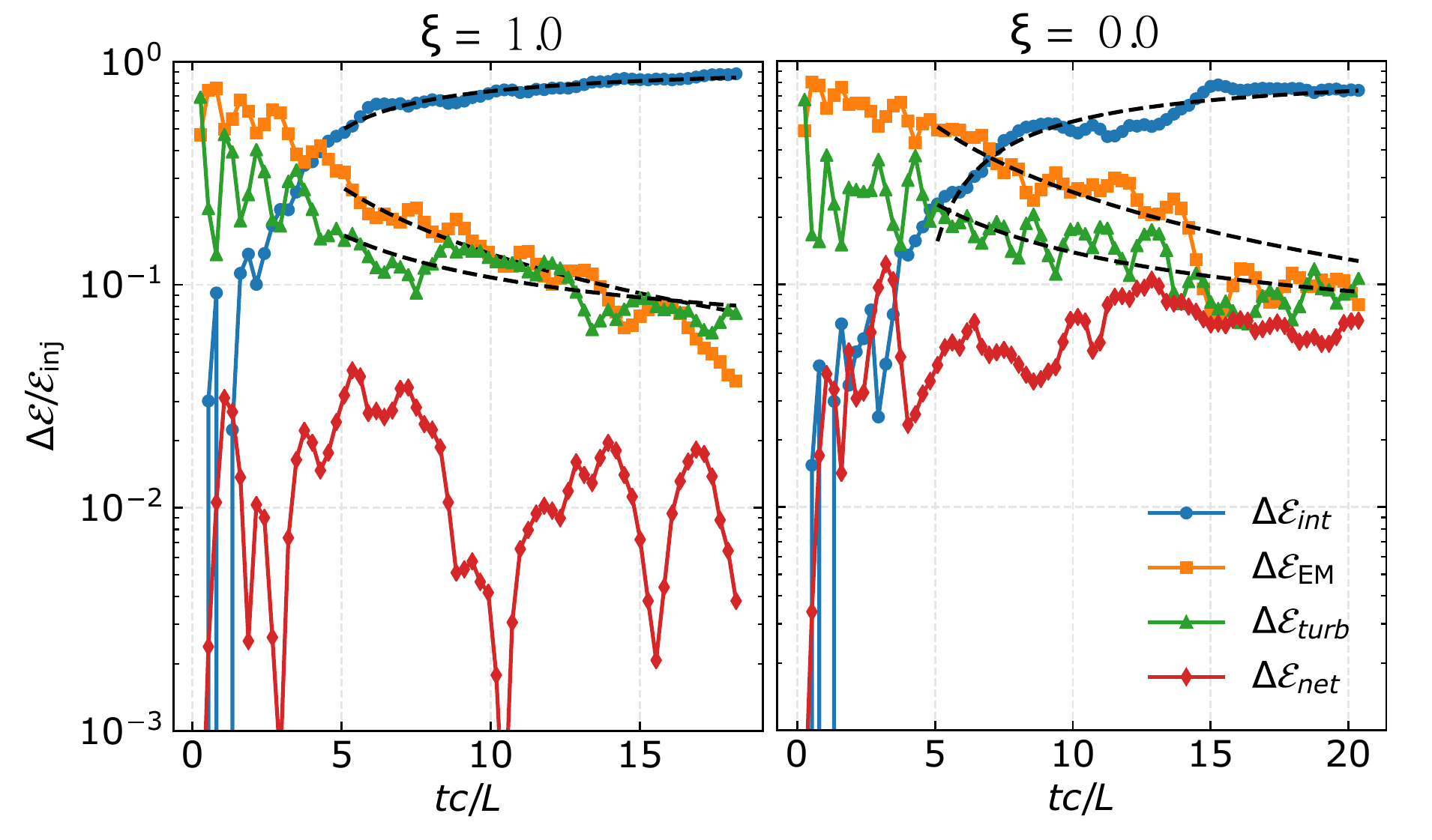}
    \caption{Time evolution of the energy partition for the balanced case ($\xi=1.0$; left) and most imbalanced case ($\xi=0.0$; right). Both show a decay in turbulent electromagnetic and kinetic energies and a saturation of internal and net flow energy densities; black dashed lines shows fits to~$A+B/t$, with~$A$ and~$B$ constants. The imbalanced case has net flow energy density about an order of magnitude higher than the balanced case and correspondingly lower internal energy density, whereas the turbulent electromagnetic and kinetic energy densities are comparable for both balance parameters. }
    \label{fig:fullPartition}
\end{figure}

Normalizing the change in each type of energy to the injected energy (the energy efficiency; Equation~\ref{eq:defEfficiency}) allows direct comparison between turbulence with different balance parameters while accounting for the injected energy's dependence on~$\xi$ (Fig.~\ref{fig:fullPartition}). By the end of the simulations at~$t=20~L/c$, the percentage of injected energy that dissipates into internal energy depends on balance parameter, varying from~$\approx80\%$ for the most imbalanced case ($\xi=0$) to~$\approx90\%$ for the balanced ($\xi=1$) case. For all balance parameters, this percentage increases over time as the majority of injected energy converts into internal energy. In contrast, the fractions of injected energy that convert into turbulent electromagnetic and kinetic energy (i.e. the electromagnetic and kinetic energy efficiencies) decrease in time as~$1/t$, plotted as dashed black lines in Fig.~\ref{fig:fullPartition}, with similar magnitudes for the balanced ($\xi=1.0$) and most imbalanced ($\xi=0.0$) case. These fits are motivated by the discussion in the previous paragraph; when normalized to the injected energy~$\propto t$, these two types of turbulent energy can be fit to the function~$A+B/t$. The fraction of injected energy that converts into net flow energy differs by an order of magnitude between the balanced case (1\%) and the imbalanced case (10\%). In both cases, however, the net flow efficiency remains relatively constant in time, indicating that a constant fraction of injected energy converts into net flow energy, with a clear dependence on~$\xi$ (also seen in Fig.~\ref{fig:grid}h). 

With a broad framework for the temporal evolution of internal, turbulent kinetic, and magnetic energy in hand, the following sections will explore each type of energy's dependence on balance parameter using various averages and highlighting statistical variation with the random seed study.

\subsubsection{Electromagnetic and Turbulent Kinetic Energy} \label{sssec:constants}
\begin{figure*}
    \centering
    \includegraphics[width=0.9\textwidth]{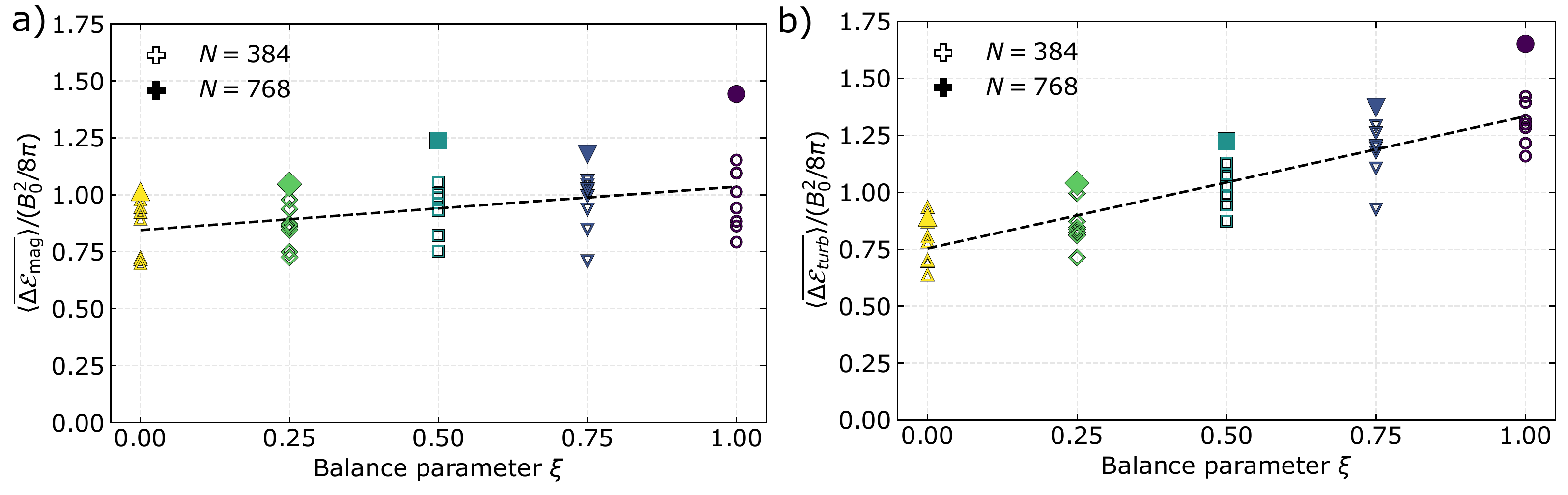}
    \caption{Trends of the turbulent magnetic and kinetic energy densities with balance parameter. Quantities are time-averaged from~$10<tc/L<20$. The largest domain size~$L/2\pi\rho_{e0}=81.5$ (filled markers) shows a linear trend with balance parameter for the turbulent magnetic (a) and kinetic (b) energy densities, respectively. The statistical deviation is shown by the~$L/2\pi\rho_{e0}=40.7$ seed study (unfilled markers). The dashed lines show linear fits. Colors and markers are the same as in Fig.~\ref{fig:grid}.}
    \label{fig:emturb-imbalance}
\end{figure*}%

After an initial transient period, the turbulent electromagnetic and kinetic energies become statistically constant in time (Fig.~\ref{fig:grid}a, b). The turbulent electromagnetic energy increases until it contains  approximately the same amount of energy as in the background field (Fig.~\ref{fig:grid}a). As $\xi$ increases from 0 to ~1, $\langle\overline{\Delta\mathcal{E}_{\rm mag}}\rangle$ increases by about 50\% for turbulence in the largest simulation domain sizes (Fig.~\ref{fig:emturb-imbalance}a).  In contrast, the electric energy density decreases from 10 - 25\% of~$B_0^2/8\pi$ for~$\xi=0.0$ to 10 - 17\% for~$\xi=1.0$ (not shown). This decrease in electric energy density could be due to the decrease in the plasma velocity, which is approximately the Alfv\'en speed:~$E\sim (\delta v/c)\times B\sim (v_A/c)B_0$. Faster heating in the balanced turbulence case leads to smaller~$v_A$ and hence smaller electric field. The simulations show that the time-averaged turbulent kinetic energy density~$\langle\overline{\Delta\mathcal{E}_{\rm turb}}\rangle$ also depends on~$\xi$, increasing from~75\% of the background magnetic energy for the most imbalanced case ($\xi=0.0$) to about~125\% for balanced turbulence with~$\xi=1.0$ (Fig.~\ref{fig:emturb-imbalance}b). The smaller values of~$\langle\overline{\Delta\mathcal{E}_{\rm mag}}\rangle$ and~$\langle\overline{\Delta\mathcal{E}_{\rm turb}}\rangle$ in the imbalanced case result from the injection of less energy in this case (Fig.~\ref{fig:einj-evolution}). The fraction of injected energy that converts into turbulent and magnetic energy (i.e. the corresponding energy efficiencies) at any given time is almost independent of balance parameter (Fig.~\ref{fig:grid}e, f), with slightly higher~$\langle\Delta\mathcal{E}_{\rm EM}\rangle/\langle\mathcal{E}_{\rm inj}\rangle$ for~$\xi=0$ compared to~$\xi=1$.
 
To test whether the turbulence is Alfv\'enic, we use the ``Alfv\'en ratio"~$r_A\equiv\langle\overline{\Delta\mathcal{E}_{\rm turb}}\rangle/\langle\overline{\Delta\mathcal{E}_{\rm mag}}\rangle$. The Alfv\'en ratio is related to the residual energy~$E_r$ (defined as the difference between the turbulent kinetic and turbulent magnetic energies) via~$E_r=(r_A-1)/(r_A+1)$. Ideal MHD predicts that the time- and volume-averaged kinetic and perturbed magnetic energies in an Alfv\'en wave (and thus perfectly imbalanced turbulence) should be in equipartition:~$r_A=1$. We might expect the same Alfv\'en ratio for turbulence (both balanced and imbalanced) comprising many Alfv\'en waves --- though due to an increase in nonlinear interactions, physical plasmas such as the solar wind often have an excess of magnetic energy such that~$r_A\approx0.7$~\citep{Chen2013a, Chen2016}; see~\citet{Boldyrev2009},~\citet{Boldyrev2011}, and~\citet{Wang2011} for MHD models of this excess. Our simulations show that both balanced and imbalanced turbulence are in equipartition to within error bars (Table~\ref{tab:magturb-ratio}). The standard deviations of mean values for~$\langle\overline{\Delta\mathcal{E}_{\rm turb}}\rangle$ and~$\langle\overline{\Delta\mathcal{E}_{\rm mag}}\rangle$ in Table~\ref{tab:magturb-ratio} were calculated for the~$L/2\pi\rho_{e0}=81.5$ and~$L/2\pi\rho_{e0}=40.7$ simulations (nine values for each balance parameter) and summed in quadrature. These error bars suggest that the large-scale ratio of turbulent kinetic to magnetic energies is independent of~$\xi$. However, the residual energy may have a scale-dependent power-law spectrum with significant dependence on imbalance, which we do not address here. The solar wind shows a clear dependence of the residual energy spectrum's slope on imbalance, with a value of~$-2$ for balanced turbulence and closer to~$-1.8$ for totally imbalanced turbulence~\citep{Chen2013a}. The dependence of the residual energy spectrum's slope on imbalance is a major outstanding puzzle that has not been successfully addressed by any phenomenological model of imbalanced turbulence thus far.

Our finding of approximate equipartition indicates that the turbulence is predominantly Alfv\'enic. In addition to Alfv\'en waves, slow and fast compressive modes also contribute to the turbulence. The fast modes, introduced by the OLA driving~\citep{Zhdankin2021} or nonlinear relativistic wave conversion~\citep{Takamoto2016}, and the slow modes, passively mixed by the turbulence~\citep{Lithwick2001}, lead to total density fluctuations on the order of 20-30\% of the background density in our simulations (not shown). Though the presence of fast and slow modes could affect the Alfv\'en ratio, characterizing their contribution is beyond the scope of this study.

\begin{table}
\centering
\caption{The turbulence in all simulations of balanced and imbalanced turbulence is approximately Alfv\'enic. The Alfv\'en ratio~$r_A=\langle\overline{\Delta\mathcal{E}_{\rm turb}}\rangle/\langle\overline{\Delta\mathcal{E}_{\rm mag}}\rangle$ is approximately 1 for the largest simulations ($L/2\pi\rho_{e0}=81.5$) for all values of the balance parameter~$\xi$. Standard deviations are calculated from the statistical seed studies at each balance parameter.}\label{tab:magturb-ratio}
\begin{tabular}{cc}
~$\xi$ &~$\langle\overline{\Delta\mathcal{E}_{\rm turb}}\rangle/\langle\overline{\Delta\mathcal{E}_{\rm mag}}\rangle$\\ \hline\hline
1.0 &~$1.1\pm0.3$\\ \hline
0.75 &~$1.2\pm0.2$ \\ \hline
0.5 &~$1.0\pm0.1$ \\ \hline
0.25 &~$1.0\pm0.1$ \\ \hline
0.0 &~$0.9\pm0.1$ 
\end{tabular}
\end{table}

\subsubsection{Internal Energy} \label{sssec:int}
The increase in internal energy dominates the plasma energy budget at late times. Though the initial internal energy starts out at about three times the initial magnetic energy for all balance parameters, the plasmas with balanced turbulence heat up almost twice as quickly as the plasmas with imbalanced turbulence (Fig.~\ref{fig:grid}c). At late times, about~$80\%$ of the injected energy is converted into internal energy in the imbalanced case ($\xi=0$), slightly lower than the corresponding value of closer to~$90\%$ for the balanced case (Fig.~\ref{fig:grid}g). 

\begin{figure}
    \centering
    \includegraphics[width=0.45\textwidth]{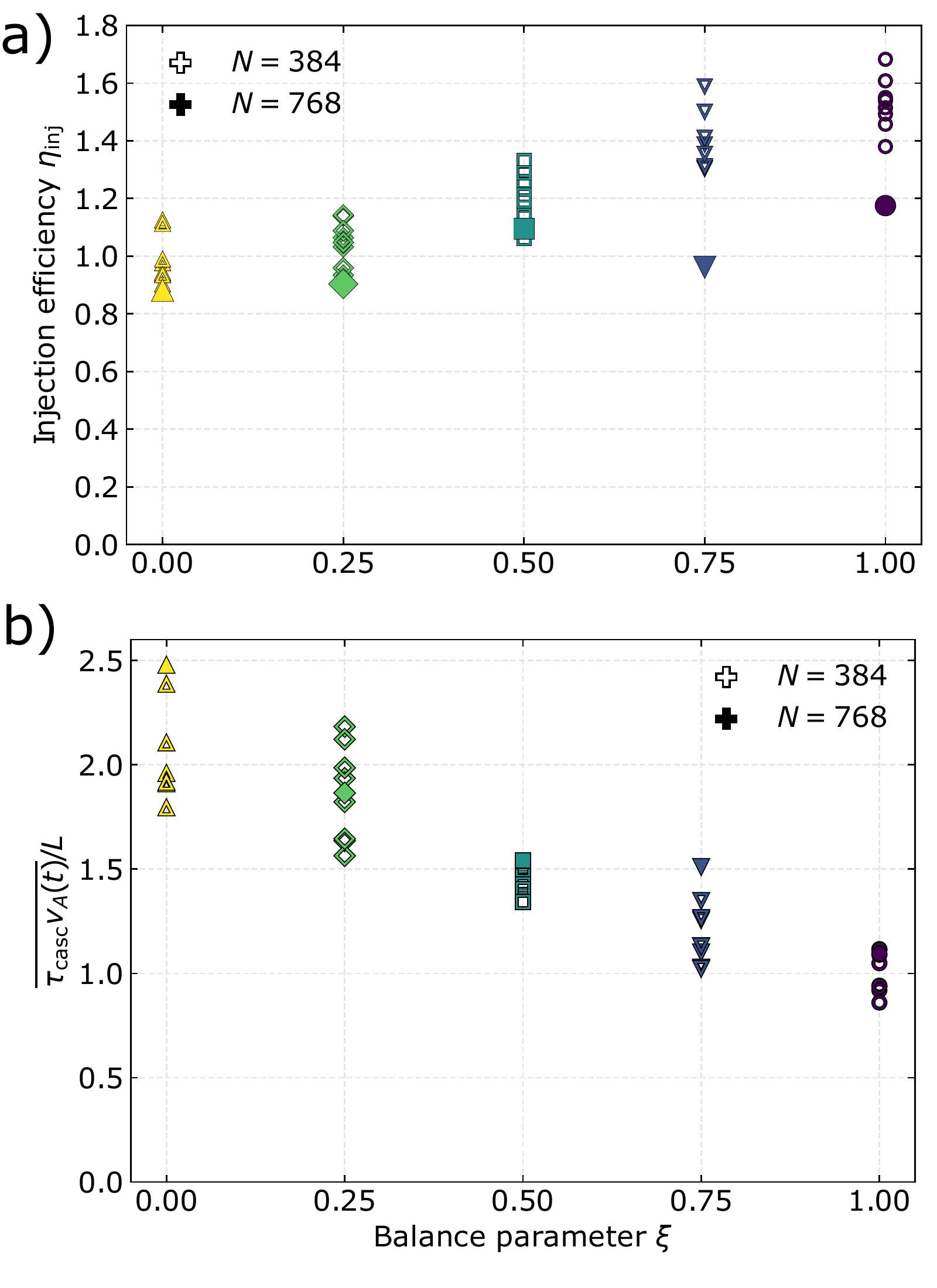}
    \caption{The injection efficiency~$\eta_{\rm inj}$~(a; Equation~\ref{eq:eta-inj}) and cascade time~$\tau_{\rm casc}$ (b; Equation~\ref{eq:tcasc}) depend on balance parameter. The largest domain size~$L/2\pi\rho_{e0}=81.5$ is shown with filled markers and the statistical deviation is shown by the~$L/2\pi\rho_{e0}=40.7$ seed study (unfilled markers). Colors and markers are the same as in Fig.~\ref{fig:grid}.}
    \label{fig:injectionEfficiency}
\end{figure}

Because the plasma's internal energy increases linearly after an initial transient, we characterize its heating rate (i.e. slope) through the dimensionless ``injection efficiency" parameter~$\eta_{\rm inj}$. We define this order-unity coefficient as the ratio of the plasma heating rate~$\langle\mathcal{\dot E}_{\rm int}\rangle(t)$ to a ``reference" heating rate~$\langle\mathcal{\dot E}_{\rm ref}\rangle$. We define~$\langle\mathcal{\dot E}_{\rm ref}\rangle$ by dividing the turbulent magnetic energy density,~$\delta B_{\rm rms}^2/8\pi$, by a characteristic nonlinear cascade time at the outer scale~$L/\delta v_{\rm rms} \sim L/(v_{A0} \delta B_{\rm rms}/B_0)$, assuming strong Alfv\'{e}nic turbulence and~$v_{A0}/c \ll 1$. This formulation gives us an operational definition of the injection efficiency in terms of quantities that can be directly measured in our simulations at any moment of time: 
\begin{equation}
    \eta_{\rm inj}\equiv\frac{\langle \mathcal{\dot E_{\rm int}}\rangle}{\langle \mathcal{\dot E_{\rm ref}}\rangle}=\frac{8\pi B_0L\langle \mathcal{\dot E_{\rm int}}\rangle}{\delta B_{\rm rms}^3 v_{A0}}. \label{eq:eta-inj}
\end{equation}
The injection efficiency quantifies how efficiently turbulent magnetic energy cascades to small scales and dissipates. The heating rate of the plasma is extracted by fitting the slope of the internal energy in the interval~$5<tc/L<20$ and converting to the injection efficiency via Equation~\ref{eq:eta-inj}, taking the value of~$\delta B_{\rm rms}$ as the time-average over the same time period. The statistical mean value of the injection efficiency varies from about 1.0 for the most imbalanced case ($\xi=0.0$) to about 1.5 for the balanced case ($\xi=1.0$) as seen in Fig.~\ref{fig:injectionEfficiency}a, with a statistical standard deviation on the order of 0.2. 

As discussed in Section~\ref{sssec:constants}, the magnitude of the magnetic field fluctuations depends on balance parameter, which in principle may influence energy dissipation and the injection efficiency. To verify that the trend in the injection efficiency~$\eta_{\rm inj}$ with balance parameter~$\xi$ is not due to the variation of the amplitude of magnetic energy perturbations~$\delta B_{\rm rms}^2/B_0^2$ with~$\xi$, we ran a simulation of imbalanced turbulence with~$\xi=0$ and a driving amplitude~$\sqrt{2}$ times its canonical value. The magnetic field fluctuation level~$\delta B_{\rm rms}^2$ changes from~$0.8 B_0^2$ for the unadjusted amplitude case to~$1.2 B_0^2$ for the increased amplitude case, consistent with the unadjusted amplitude of the balanced case with the same random seed. The injection efficiency, however, increases to~1.1 with the increased amplitude, compared to 0.9 for the unadjusted~$\xi=0$ case and 1.4 for the unadjusted~$\xi=1$ case with the same random seed. Thus, two simulations with the same level of magnetic field perturbations but different balance parameters experience significantly different injection efficiencies, suggesting that the injection efficiency has an inherent dependence on balance parameter rather than~$\delta B_{\rm rms}$. This result suggests that the cascade time depends on balance parameter.

To explore the possibility that the cascade time~$\tau_{\rm casc}$ depends on balance parameter, we define 
\begin{equation}
    \tau_{\rm casc}\equiv \frac{\langle\Delta\mathcal{E}_{\rm EM} + \mathcal{E}_{\rm turb}\rangle}{\langle\dot{\mathcal{E}}_{\rm int}\rangle}. \label{eq:tcasc}
\end{equation}
The cascade time is normalized to the global Alfv\'en time~$L/v_A(t)$ and then time-averaged over~$10<tc/L<20$ (i.e.~$5.8<tv_{A0}/L<11.6$). We expect the cascade time to be on the order of an Alfv\'en time for an Alfv\'enic cascade, and for balanced turbulence we indeed find that the cascade time varies statistically between~$0.8$ and~$1.2~L/v_A(t)$ (Fig.~\ref{fig:injectionEfficiency}b). However, the cascade time increases to about~$2.2~L/v_A(t)$ on average for the imbalanced case~$\xi=0$. The lack of overlap between the cascade times of imbalanced and balanced turbulence suggest that the difference is statistically significant, rather than a fluke of random seeds. A longer cascade time for more imbalanced turbulence is consistent with~\citet{Lithwick2007}'s suggestion that the dominant waves are less strongly scattered in the imbalanced case. 

\subsection{Net Flow Energy and Momentum Transfer}\label{ssec:net}
As a component of the energy not present in balanced turbulence, we expect the kinetic energy in the net motion of the plasma through the simulation domain to depend on the balance parameter. In the balanced case, the statistically-symmetric (although not necessarily momentum-conserving) driving should on average lead to no net motion. In contrast, because imbalanced driving breaks the symmetry along the background magnetic field, we may expect a nonzero net flow energy for~$\xi<1.0$ if the asymmetric wave momentum converts into plasma momentum. In a gravitational potential, the net flow that results from efficient wave-plasma momentum coupling could form a wind or outflow. In this section, we first propose a simple model for the properties of such a net flow (Section~\ref{sssec:netanalytic}) and then compare to numerical results (Section~\ref{sssec:netnumerics}).

\subsubsection{Analytic Framework for Momentum Transfer}\label{sssec:netanalytic}
A net flow could result from the efficient transfer of injected wave momentum into plasma momentum. In this section, we propose that the net flow velocity should be constant and provide a scaling for its magnitude. We can write the net flow energy density as
\begin{align}
    \mathcal{E}_{\rm net}(t)&=\left(\Gamma_{\rm net}(t)-1\right)\langle \rho\rangle(t)c^2 \approx\frac{\langle\mathcal{E}_{\rm int}\rangle(t) v_{\rm net}^2(t)}{2c^2},
\end{align}
where the last expression holds for~$v_{\rm net}(t)\ll c$. We have defined the net Lorentz factor $\Gamma_{\rm net}(t)=(1-v_{\rm net}^2/c^2)^{-1/2}$.

Analogously, the internal energy density relates to the net plasma momentum density as
\begin{equation}
    \langle\mathcal{P}_{\rm z,tot}\rangle(t)=\Gamma_{\rm net}(t)\langle\rho\rangle(t) v_{\rm net}(t)\approx\frac{\langle\mathcal{E}_{\rm int}\rangle(t)}{c^2} v_{\rm net}(t), \label{eq:vz}
\end{equation}
where again~$v_{\rm net}\ll c$ in the last expression. In simulations, both~$\mathcal{E}_{\rm net}(t)$ and~$\langle\mathcal{P}_{\rm z,tot}\rangle(t)$ increase linearly in time (Section~\ref{sssec:netnumerics}). Because these two quantities depend on different powers of~$v_{\rm net}$, we deduce that the net velocity should be relatively constant in time. We test this prediction in the next section.

We can understand the net flow as a relativistic effect. As a limiting case, assume that the maximal asymmetric Poynting flux for a single Alfv\'en wave $|\langle S_z\rangle_{1-\rm wave}|$ (Equation~\ref{eq:maxSz}) is injected into a plasma. If a fraction~$\epsilon$ of the momentum density~$|\langle S_z\rangle_{1-\rm wave}|/c^2$ in this electromagnetic wave converts into the plasma momentum density~$\langle\mathcal{P}_{\rm z,tot}\rangle$, we have
\begin{equation}
    \langle\mathcal{P}_{\rm z,tot}\rangle=\epsilon\frac{\langle \delta B^2\rangle v_{A0}}{4\pi c^2}=\Gamma_{\rm net}\langle\rho\rangle v_{\rm net},
\end{equation} 
where the last equality follows from Equation~\ref{eq:vz}. By writing the relativistic mass density~$\langle\rho\rangle=\langle\gamma\rangle m_en_0$, we find
\begin{align}
    \Gamma_{\rm net}v_{\rm net}&=\frac{\langle\delta B^2\rangle}{4\pi \langle\rho\rangle c^2}\epsilon v_{A0}=\frac43\epsilon~\delta\sigma~v_{A0}\\
    v_{\rm net}&=c\epsilon\sqrt{\frac{16~\delta\sigma^2~ v_{A0}^2/c^2}{9+16~\delta\sigma^2~v_{A0}^2/c^2}}\to \frac43\epsilon~\delta\sigma~v_{A0}\label{eq:vnetSigma}
\end{align}
where we have defined~$\delta\sigma=\langle\delta B^2\rangle/(4\pi h)$ and the last expression again holds for~$v_{\rm net}\ll c$. From Equation~\ref{eq:vnetSigma}, we see that the net flow's velocity approaches 0 as the magnetization goes to~0. Though the efficiency~$\epsilon$ of converting electromagnetic momentum into plasma momentum can never be greater than 1, it could change with~$\sigma$.

\subsubsection{Numerical Results for Momentum Transfer}\label{sssec:netnumerics}
\begin{figure}
    \centering
    \includegraphics[width=0.45\textwidth]{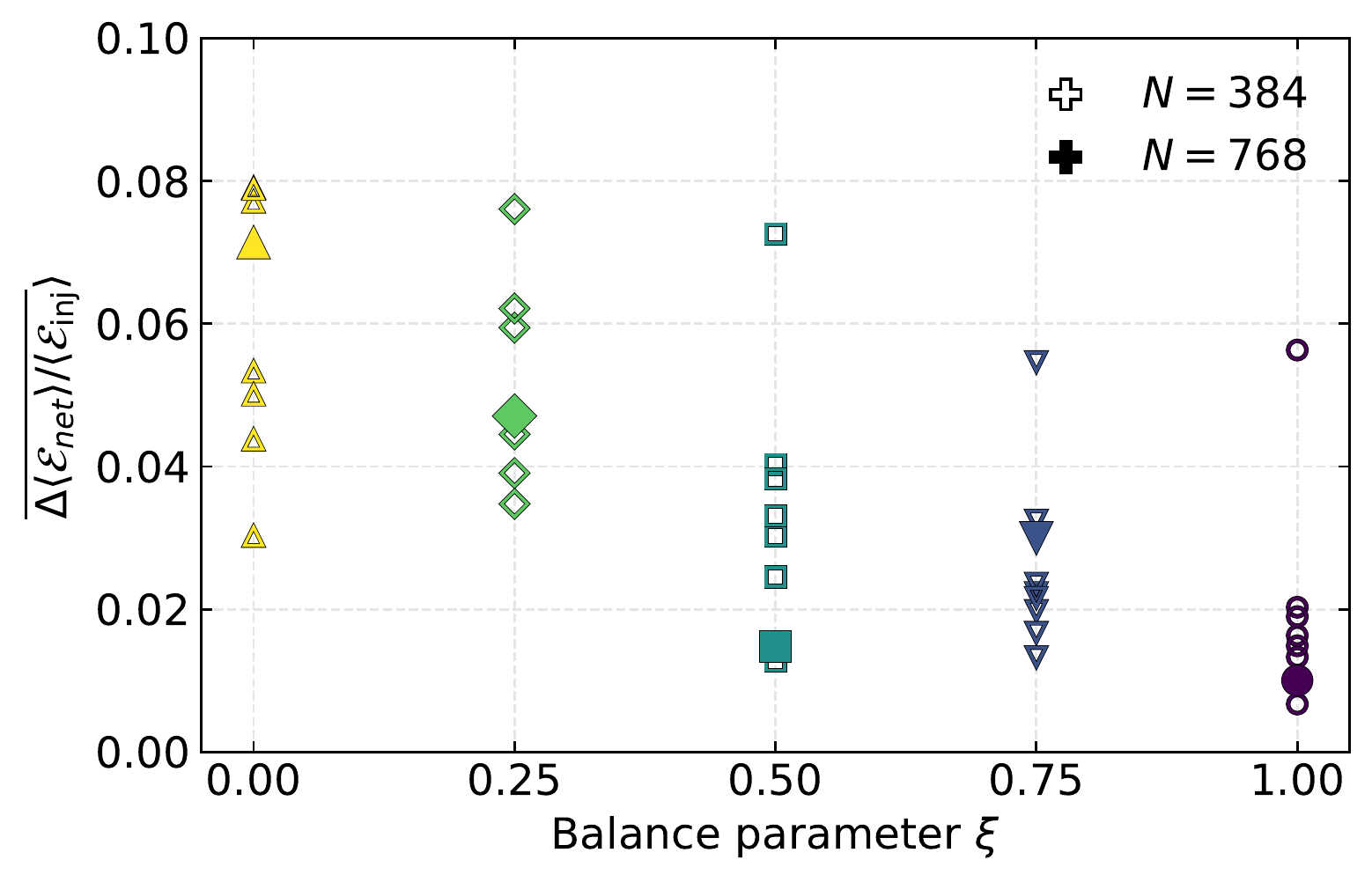}
    \caption{The net flow energy efficiency decreases with increasing balance parameter. The plotted values are volume-averaged and time-averaged from~$10<tc/L<20$. The largest domain size~$L/2\pi\rho_{e0}=81.5$ is shown with filled markers and the statistical deviation is shown by the~$L/2\pi\rho_{e0}=40.7$ seed study (unfilled markers). Note that the outliers for~$\xi=0.5$, 0.75, and 1.0 with net flow energy efficiencies a factor of 2 higher than the rest of the seed study were run with the same random seed. Colors and markers are the same as in Fig.~\ref{fig:grid}.}
    \label{fig:net-scaling}
\end{figure}%
\begin{figure*}
    \centering
    \includegraphics[width=0.8\textwidth]{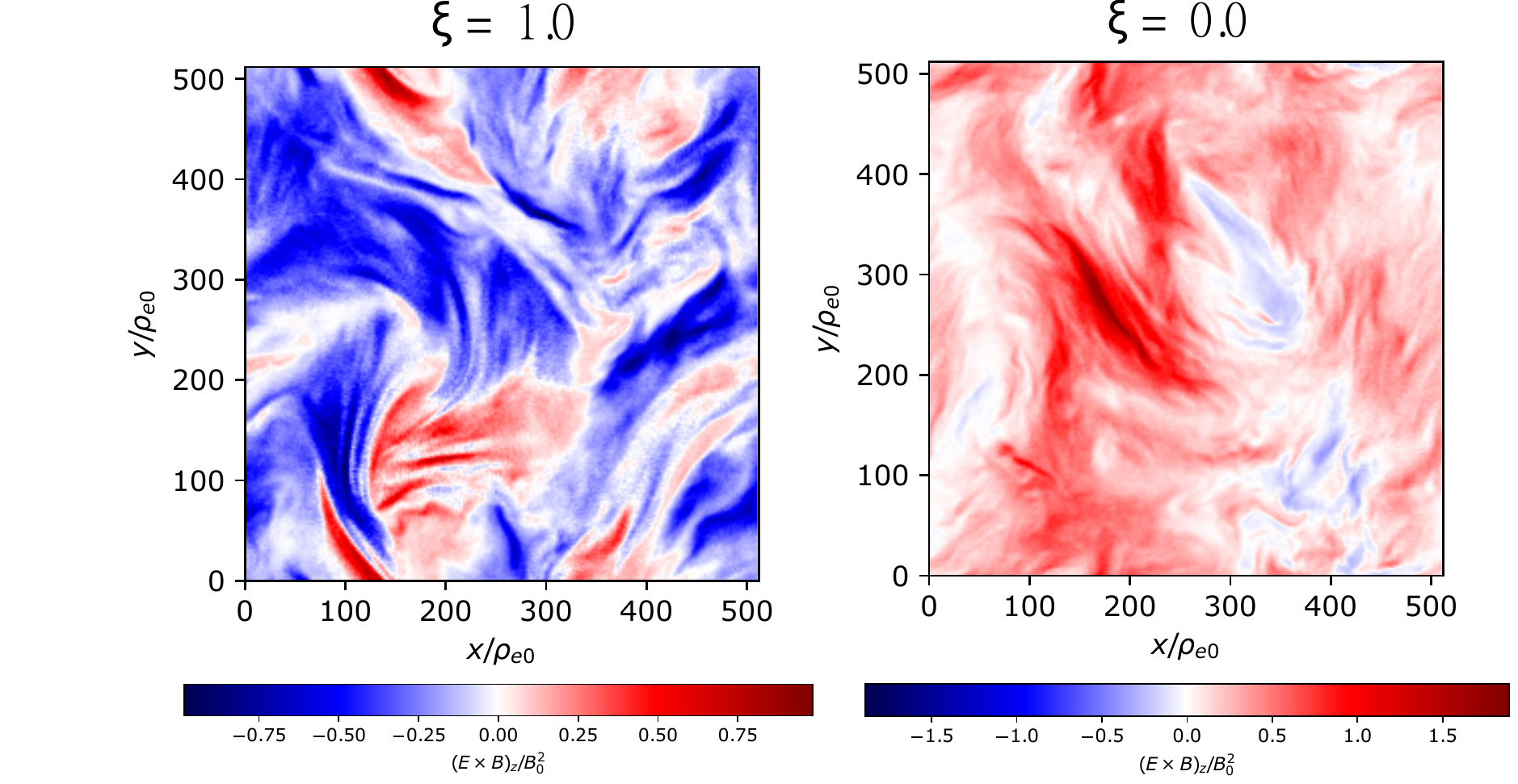}
    \caption{Even turbulence that is balanced as a whole has spatial and temporal pockets of locally imbalanced turbulence. Slices of the Poynting flux~$(c/4\pi){\boldsymbol E}\times{\boldsymbol B}$ in the~$z-$direction, taken at the plane~$z=0$ at time~$t=16.1~L/c$ and normalized to~$(c/4\pi)B_0^2$ for balanced turbulence ($\xi=1.0$; left) and most imbalanced turbulence ($\xi=0.0$; right) show variation in the sign of Poynting flux throughout the domain.}
    \label{fig:ExBz-slice}
\end{figure*}%
In this section, we test the assumptions behind the above calculations and demonstrate that our simulations do indeed find a net flow in line with the above framework.

Our simulations find that about~$8\%$ of the injected energy converts into net flow energy even in the most imbalanced case ($\xi=0.0$), a fraction that becomes comparable to the turbulent electromagnetic or kinetic energy efficiencies after the latter two have decayed by a factor of two or more, around~$15~L/c$ (Fig.~\ref{fig:fullPartition}). Since the net flow efficiency fluctuates strongly about a mean value in time, we compare different balance parameters by averaging over~$10<tc/L<20$. The result, shown in Fig.~\ref{fig:net-scaling}, reveals that the net flow efficiency increases with decreasing~$\xi$, as expected. The mean over an ensemble of identical simulations with~$\xi=0$ and varying random seeds is about 0.06, approximately three times as large as that for the balanced turbulence (about~0.02). The statistical spread in the balanced case is on the order of~0.01, and about~0.03 for the imbalanced case. Because energy is a strictly positive quantity and a limited number of modes were driven, even the simulations of balanced turbulence ($\xi=1.0$) have a non-zero (albeit small) net flow energy due to short periods of net motion. Notably, there are three prominent outliers in Fig.~\ref{fig:net-scaling} in the magnitude of the net flow efficiency in simulations with balance parameters~$\xi=0.5$,~0.75, and~1.0. Each of these outlying simulations was initialized with the same random seed, suggesting that the particular random phases of the driven modes resulted in non-zero mean velocity later in the simulations' evolutions. The presence of these outliers demonstrates the need for a suite of random seeds to tease out statistically-robust trends. 
\begin{figure*}
    \centering
    \includegraphics[width=0.9\textwidth]{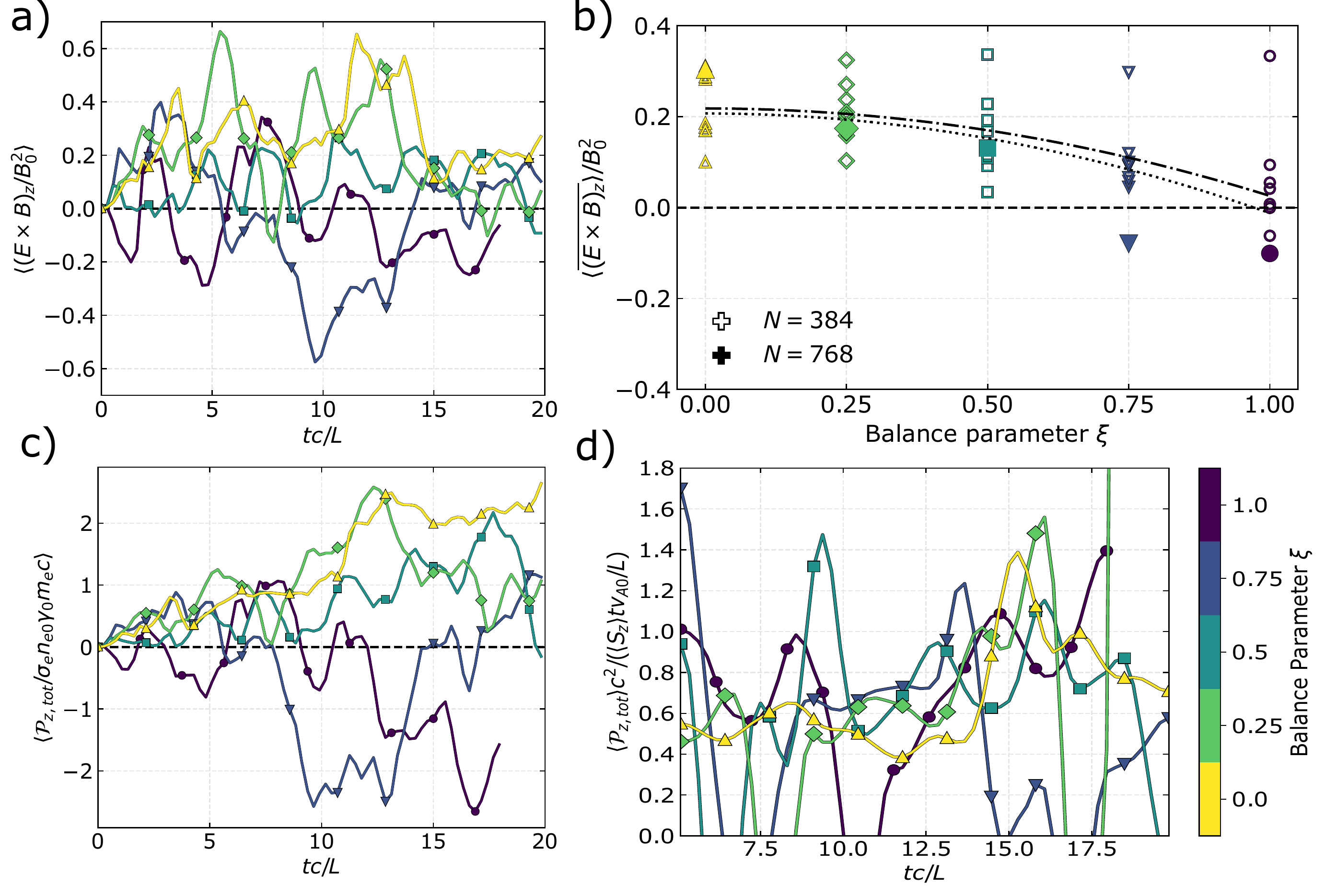}
    \caption{The average parallel Poynting flux is approximately constant in time, whereas the~$z-$momentum of the plasma increases linearly in time. a) The time evolution of the volume-averaged Poynting flux~$(c/4\pi)\langle{\boldsymbol E}\times{\boldsymbol B}\rangle$ in the~$z-$direction, normalized to~$(c/4\pi)B_0^2$, shows fluctuations around some mean value; time-averaging the curves over~$10<tc/L<20$ shows a quadratic dependence on balance parameter (b). The time evolution (c) of the parallel plasma momentum shows an increase in time. The ratio~$\langle\mathcal{P}_{z, {\rm tot}}c^2\rangle/\left(\langle S_z\rangle~tv_{A0}/L\right)$, shown in (d), is of order unity for all values of balance parameter. Black dash-dot lines show a quadratic fit; the dotted line is a quadratic fit without the outlier seed. Dashed black lines indicate zero. Colors and markers are the same as in Fig.~\ref{fig:grid}.}
    \label{fig:momentum}
\end{figure*}%
\begin{figure*}
    \centering
    \includegraphics[width=0.9\textwidth]{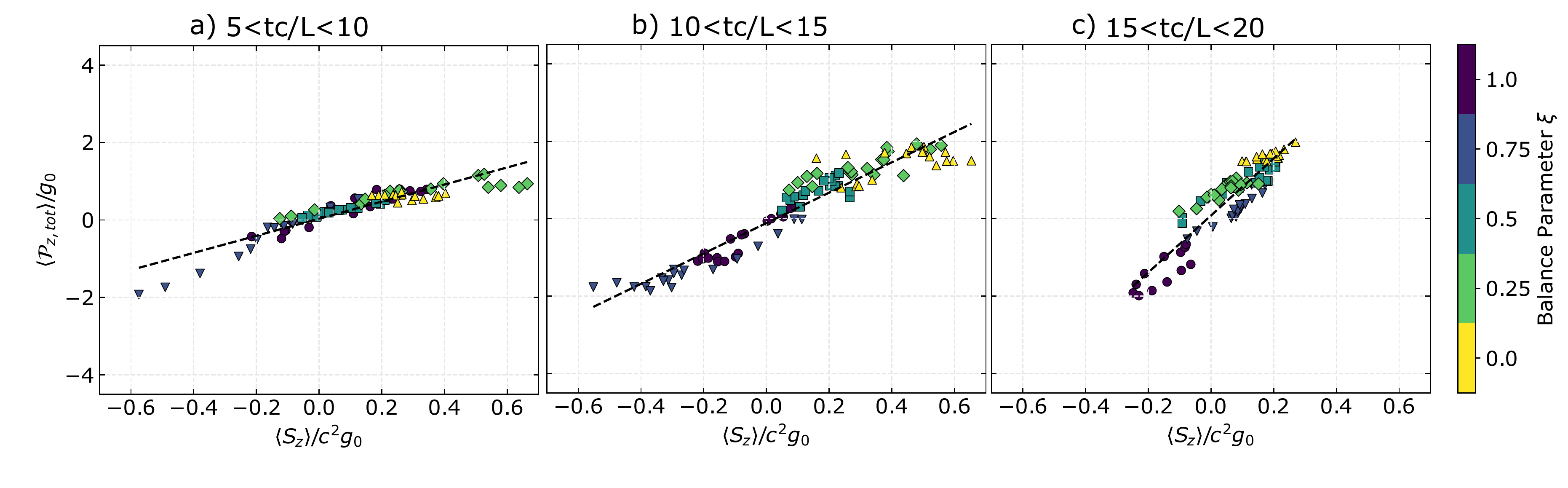}
    \caption{The parallel electromagnetic momentum generates and maintains the parallel momentum of the plasma. In these plots, each point corresponds to the instantaneous values of the volume-averaged plasma momentum along the background magnetic field and the volume-averaged parallel electromagnetic momentum at a given simulation time. Three periods of time are shown with linear fits of slopes 2.2, 3.9, and 7.4 for (a)~$5<tc/L<10$, (b)~$10<tc/L<15$, and (c)~$15<tc/L<20$, respectively. Here,~$g_0=B_0^2/(4\pi c)$ is a typical value of momentum density. Colors and markers are the same as in Fig.~\ref{fig:grid}.}
    \label{fig:scatter}
\end{figure*}%
\begin{figure*}
    \centering
    \includegraphics[width=0.9\textwidth]{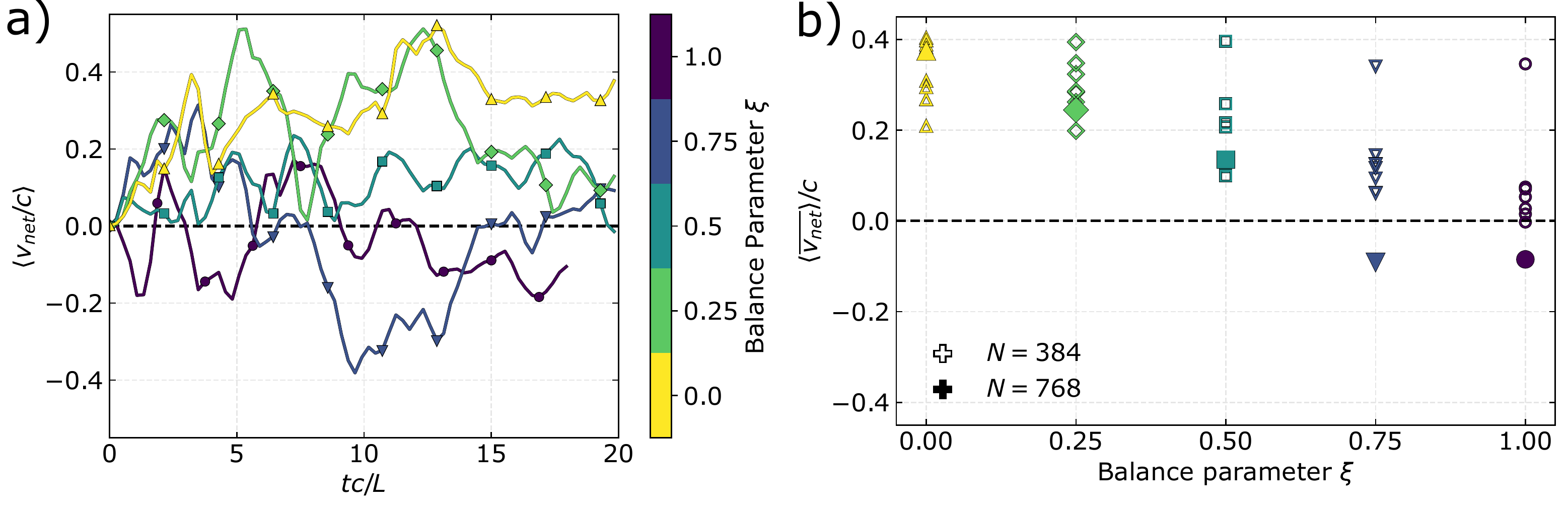}
    \caption{The net plasma velocity along the background magnetic field depends on balance parameter. a) The time evolution of the volume-averaged velocity~$v_{\rm net}$ (Equation~\ref{eq:vz}) shows values that fluctuate in time around some mean value that depends on~$\xi$; time-averaging the curves over~$10<tc/L<20$ shows a dependence on balance parameter (b). The largest domain size~$L/2\pi\rho_{e0}=81.5$ is shown with filled markers and the statistical deviation is shown by the~$L/2\pi\rho_{e0}=40.7$ seed study (unfilled markers). Colors and markers are the same as in Fig.~\ref{fig:grid}. Black dashed lines show zero.}
    \label{fig:vz}
\end{figure*}%

Although the net flow energy density demonstrates the importance of the net flow in the overall energy budget, it does not contain information about the direction of the plasma's net motion. To address this issue, we now examine momentum rather than energy. First, we discuss the injected electromagnetic momentum and look at the Poynting flux~${\boldsymbol S}=(c/4\pi){\boldsymbol E}\times{\boldsymbol B}$. For balance parameters~$\xi<1.0$, we expect the driven waves' Poynting flux to be nonzero along the background magnetic field (``parallel Poynting flux";~$S_z$). The spatial distribution of the parallel Poynting flux, plotted in Fig.~\ref{fig:ExBz-slice}, is highly nonuniform. Similar to MHD turbulence~\citep{Perez2009}, our balanced simulation has local patches of strong Poynting flux and thus high imbalance, highlighting the fundamental connection between balanced and imbalanced turbulence at small scales. The total, volume-averaged parallel Poynting flux~$\langle S_z\rangle$ is statistically constant in time and, for the balanced case ($\xi=1.0$), oscillates around zero (Fig.~\ref{fig:momentum}a). After time-averaging over the period from~$10<tc/L<20$, the parallel Poynting flux is clearly positive for imbalanced turbulence and consistent with zero for balanced turbulence (Fig.~\ref{fig:momentum}b). The value~$\langle\overline{S_z}\rangle\approx0.2~(cB_0^2/4\pi)$ at~$\xi=0.0$ is about 30\% of the limiting value~$(\delta B/B_0)^2 v_A/c~(c B_0^2/4\pi)\approx v_A/c~(c B_0^2/4\pi)\approx0.58~(c B_0^2/4\pi)$ expected for a single Alfv\'en wave (Equation~\ref{eq:maxSz}). The decrease in the Poynting flux with increasing balance parameter agrees well with the quadratic fit predicted by Equation~\ref{eq:szhcscaling}, as shown by the dotted and dash-dot black lines in Fig.~\ref{fig:momentum}b, though we do not rule out a linear dependence. The same outliers discussed for the net flow efficiency are present in the parallel Poynting flux. The time- and volume-averaged Poynting flux in the~$x-$ and~$y-$directions are much smaller than the parallel Poynting flux: within~$0.1~(c~B_0^2/4\pi)$ of zero for all balance parameters (not shown), indicating that the net electromagnetic momentum is primarily along the background magnetic field.

The injected Poynting flux imparts net momentum to the plasma. In agreement with the interpretation in Section~\ref{sssec:framework}, the volume-averaged parallel momentum~$\langle\mathcal{P}_{z, {\rm tot}}\rangle$ of the plasma increases approximately linearly over time (Fig.~\ref{fig:momentum}c). The ratio~$\langle\mathcal{P}_{z, {\rm tot}}c^2\rangle/\left(\langle S_z\rangle~tv_{A0}/L\right)$, shown in Fig.~\ref{fig:momentum}d, is approximately constant in time and fluctuates around values of order unity for any given balance parameter. In this figure, the parallel Poynting flux has been converted to an electromagnetic momentum density~$\langle S_z\rangle/c^2$, and the momentum densities of both the plasma and the electromagnetic waves are normalized to~$g_0\equiv B_0^2/(4\pi c)$, allowing for direct comparison between the two quantities. The ratio of the volume averages has been smoothed with a Hanning window to remove arbitrarily large values due to a small Poynting flux. To further illustrate the relationship between~$\langle S_z\rangle$ and~$\langle \mathcal{P}_{z, {\rm tot}}\rangle$, Fig.~\ref{fig:scatter} shows their values at each time snapshot (given by individual dots) for each balance parameter (shown via marker and color). In general, the plasma momentum dominates over the electromagnetic momentum, with a time-dependent ratio given by the slope of the linear fit. The positive ratio shows that a positive volume-averaged parallel electromagnetic momentum corresponds to a positive parallel plasma momentum, as expected if Poynting flux converts to plasma momentum. The balanced turbulence's electromagnetic momentum and plasma momentum span both positive and negative values; for more imbalanced turbulence, the distribution shifts up and right, demonstrating that the asymmetric driving of electromagnetic momentum results in asymmetric net motion of the plasma in the~$z$-direction. 

Using Equation~\ref{eq:vz} to solve for $v_{\rm net}$ shows that~$v_{\rm net}$ fluctuates around a mean value dependent on balance parameter (Fig.~\ref{fig:vz}a). The net velocity of the plasma with balanced turbulence oscillates around zero, never reaching more than~$0.2c$. The plasmas with the most imbalanced turbulence can reach net velocities up to~$0.5c$. Averaging over~$10<tc/L<20$ shows a clear dependence of the net velocity on balance parameter (Fig.~\ref{fig:vz}b). As expected, plasmas with balanced turbulence experience a time-averaged net velocity near zero, though the finite simulation duration means that temporary movements parallel or anti-parallel to the background magnetic field are not completely averaged out. Equation~\ref{eq:vnetSigma} predicts a net velocity of~$0.4c$ for the most imbalanced turbulence (plugging in four waves with~$\delta B\sim B_0$, and setting~$\sigma_0=0.5$,~$v_{A0}=0.58c$ and~$\epsilon=1$), remarkably close to the values found in Figure~\ref{fig:vz}b.

Previous studies of imbalanced turbulence appear to exclude the possibility of generating net plasma motion along the background magnetic field, either through the assumption of reduced MHD, gyrokinetics, or force-free description~\citep{Cho2014, Perez2012,Meyrand2020}. As such, this work presents, to our knowledge, the first numerical demonstration and investigation of net flow due to imbalanced turbulence. Such a large net motion of the plasma may have implications for driving accretion disk winds, particularly if the wind comprises mostly nonthermal particles.

\subsection{Nonthermal Particle Acceleration} \label{ssec:ntpa}
\begin{figure}
    \centering
    \includegraphics[width=0.45\textwidth]{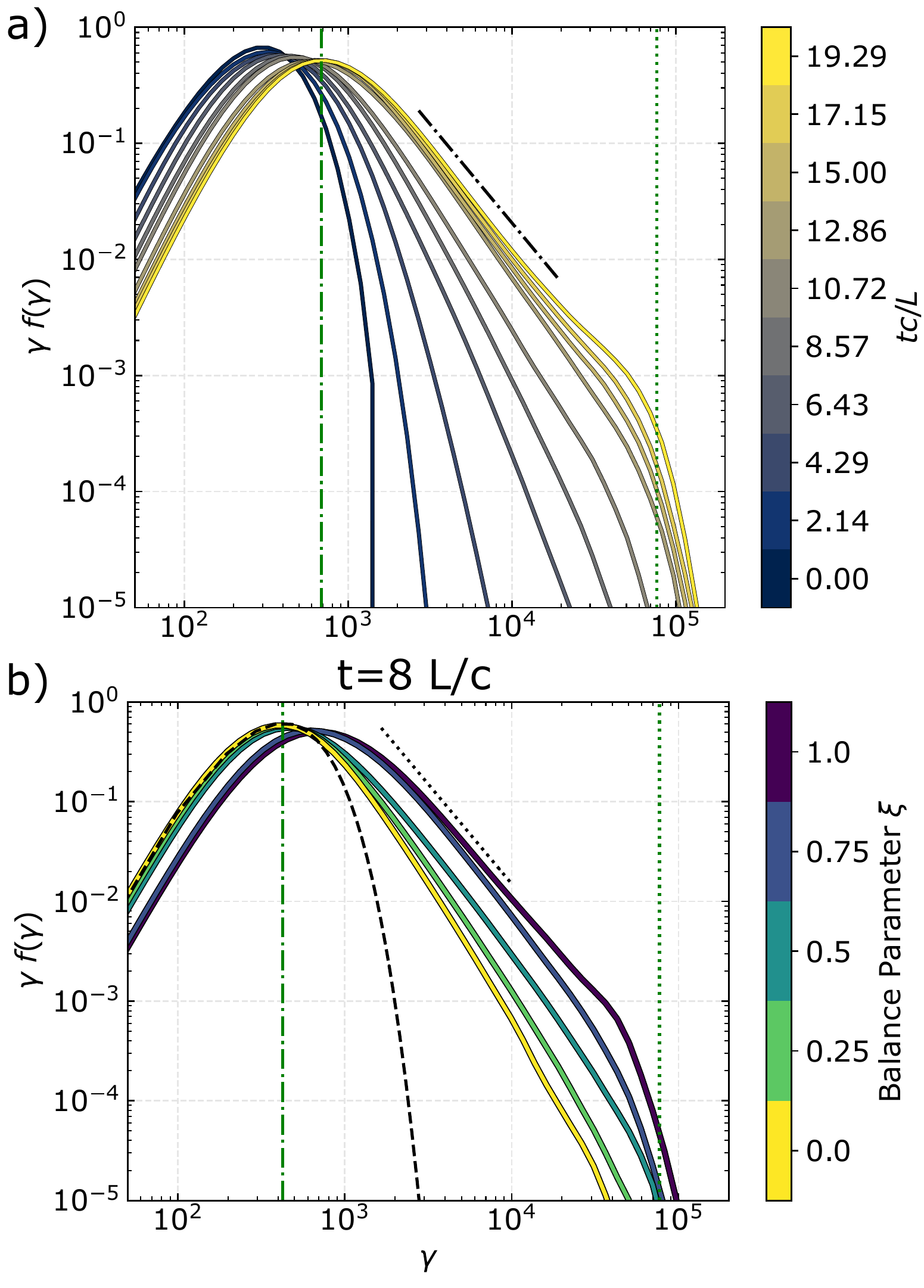}
        \caption{Particle acceleration occurs for all values of balance parameter. a) The distribution function of the most imbalanced case ($\xi=0.0$) becomes shallower in time from an initial Maxwell-J\"uttner distribution (purple) to a Maxwell-J\"uttner distribution plus a hard power-law component at later times (yellow). b) Spectra taken at the same time~$t=8.0~L/c$ for different balance parameters show different peak energies but similar power-law components. The dashed black line shows a Maxwell-J\"uttner fit to the~$\xi=0.0$ case. The vertical green dash-dot line shows the mean Lorentz factor~$\langle\gamma\rangle$ extracted from this fit. The vertical green dotted line shows the maximum energy~$\gamma_{\rm max}$. The dash-dot black line in panel (a) shows the power law~$\gamma^{-2.7}$, while the dotted black line in panel (b) shows~$\gamma^{-3}$. Colors are the same as in Fig.~\ref{fig:grid}.}
    \label{fig:particle_acceleration}
\end{figure}%
\begin{figure}
    \centering
    \includegraphics[width=0.45\textwidth]{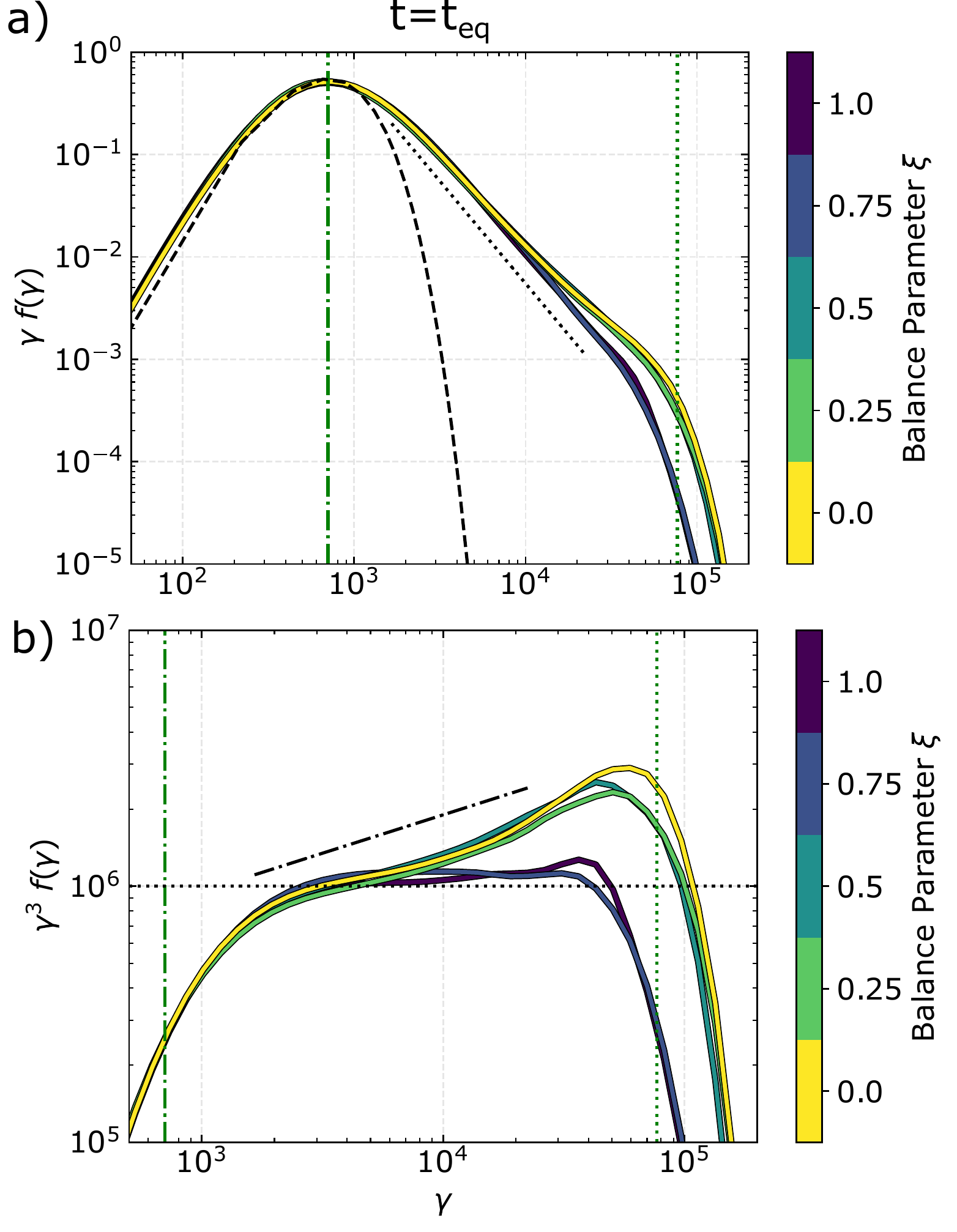}
    \caption{At equivalent times, the power laws of imbalanced turbulence are slightly flatter/harder than balanced turbulence. a) The particle energy spectra at equivalent times (see Table~\ref{tab:teq}) show a similar mean energy and similar power laws until~$\gamma\sim10^4$. The dashed black line shows a Maxwell-J\"uttner fit to the~$\xi=0.0$ case. b) Compensating by~$\gamma^3$ reveals that more imbalanced turbulence ($\xi=0.0$, 0.25, and~0.5) has flatter power laws than the more balanced turbulence with~$\xi=0.75$ or~1.0. The vertical green dash-dot line shows the mean Lorentz factor~$\langle\gamma\rangle$ extracted from the Maxwell-J\"uttner fit. The vertical green dotted line shows the maximum energy~$\gamma_{\rm max}$. The black dash-dot line in panel (b) shows the spectrum compensated to~$\gamma^3$, while the dotted black line shows the power law~$\gamma^{-3+3}$ (a constant). Colors are the same as in Fig.~\ref{fig:grid}.}
    \label{fig:particle_acceleration_teq}
\end{figure}%
Nonthermal particle acceleration can explain high-energy flares and power laws seen in spectra of various astrophysical systems. Studying NTPA self-consistently requires PIC simulations. Recent PIC simulations of turbulence~\citep{Zhdankin2017,Zhdankin2018,Comisso2018,Comisso2019} have successfully produced nonthermal particle populations that result in power-law spectra. Similar results have been produced in PIC simulations of kink-unstable jets~\citep{Alves2018,Davelaar2020} and the magnetorotational instability in accretion discs~\citep{Riquelme2012,Hoshino2013,Kunz2016}, where turbulence may play a fundamental role in the particle acceleration. Because power-law spectra are observed in systems with asymmetric energy injection, it is important to understand how imbalance affects particle acceleration.

We find that imbalanced turbulence can accelerate a significant portion of the plasma's particles to suprathermal energies (Fig.~\ref{fig:particle_acceleration}), much like balanced turbulence at similar magnetizations~$\sigma\sim1$. Even the most imbalanced case~$\xi=0.0$ shows the development of a high-energy power-law tail, which hardens and reaches an asymptotic slope after about~$12~L/c$ (Fig.~\ref{fig:particle_acceleration}a). At late times~$t\gtrsim12~L/c$, the simulation domain's boundary conditions limit the maximum attainable Lorentz factor to~$\gamma_{\rm max}=LeB_0/m_ec^2$, resulting in the ``pile-up" of high-energy particles at~$\gamma_{\rm max}$, followed by a sharp cutoff rather than the continuation of the power law to even higher energies~\citep{Zhdankin2018b}. Visually, the nonthermal distribution matches the power-law scaling~$f(\gamma)\propto \gamma^{-2.7}$ between~$\langle\gamma\rangle$ and~$\gamma_{\rm max}$, shown by the dash-dot line in Fig.~\ref{fig:particle_acceleration}a. Comparing turbulence with different balance parameters, we see that more balanced turbulence heats the plasma more quickly and forms a power-law spectrum faster than imbalanced turbulence (Fig.~\ref{fig:particle_acceleration}b). For particle spectra taken at~$t=8.0~L/c$, the simulation of balanced turbulence~$\xi=1.0$ has already heated to a peak Lorentz factor of about 700 and is experiencing pile-up, as shown by the spectrum's break around~$\gamma\sim5\times10^{4}$. In contrast, the most imbalanced case~$\xi=0.0$ still has a peak Lorentz factor of around 400 and its power-law index has not yet reached an asymptotic value (Fig.~\ref{fig:particle_acceleration}b).

\begin{table}
\centering
\caption{Table of ``equivalent" times~$t_{\rm eq}$ in~$L/c$ where the same amount of energy density~$10.3~B_0^2/8\pi$ has been injected. Values of the nonthermal particle and energy fraction at~$t_{\rm eq}$, shown in the second and third rows, are discussed in Section~\ref{ssec:ntpa}. Equivalent times are labelled in Fig.~\ref{fig:einj-evolution} as red~$\times$'s.}\label{tab:teq}
\begin{tabular}{cccccc}
$\xi$ &~$0.0$ &~$0.25$ &~$0.5$ &~$0.75$ &~$1.0$ \\ \hline\hline
$t_{\rm eq}$ & 20.0   & 17.4      & 13.1      & 9.4       & 8.0      \\
$N_{\rm nonthermal}/N_{\rm total}$ & 0.20 & 0.19 & 0.20 & 0.21 & 0.21 \\ 
$\langle\mathcal{E}_{\rm nonthermal}\rangle/\langle\mathcal{E}_{\rm pl}\rangle$ & 0.54 & 0.53 & 0.55 & 0.54 & 0.54\\
\end{tabular} 
\end{table}%
Because of the different rates at which energy is injected into simulations with different balance parameters, it may be more meaningful to compare particle spectra not at the same fixed absolute time, but rather at ``equivalent times" when a fixed amount of energy has been injected. For definiteness, we take this time to coincide with the end ($t=20~L/c$) of the simulation for the most imbalanced case,~$\xi=0$, corresponding to an injected energy of~$10.3~B_0^2/8\pi$. The equivalent times for the simulations vary from~$8~L/c$ for the balanced case~$\xi=1.0$ to~$20~L/c$ for the most imbalanced case~$\xi=0.0$ (Table~\ref{tab:teq}). When compared at these equivalent times, particle spectra for different~$\xi$ essentially collapse to a single universal curve (Fig.~\ref{fig:particle_acceleration_teq}a). In particular, the peak Lorentz factors and the power-law tails up to~$\gamma \approx 10^4$ become nearly indistinguishable. This similarity suggests that NTPA operates similarly in balanced and imbalanced turbulence when considered on appropriate timescales. In particular, the nonthermal segments of the distribution function match the power-law scaling~$f(\gamma)\propto\gamma^{-3}$ (dotted line) for all values of balance parameter, suggesting that imbalanced and balanced turbulence accelerate particles with the same asymptotic spectra. Finer differences appear at higher energies when the spectra are compensated by~$\gamma^3$; whereas the balanced cases~$\xi=1.0$ and 0.75 closely follow this~$\gamma^{-3}$ scaling in the interval~$200\lesssim\gamma\lesssim4\times10^4$, the more imbalanced cases are never quite flat and appear to more closely follow the scaling~$\gamma^{-2.7}$ (dash-dot line), as shown in Fig.~\ref{fig:particle_acceleration_teq}b. This power-law index of~$-2.7$ matches the power-law index of balanced turbulence particle spectra in smaller box sizes~\citep{Zhdankin2018b}, suggesting that pile-up contaminates the spectra. Because the equivalent times for the simulations of imbalanced turbulence are much longer, the high-energy pile-up could be due to a small sub-population of particles whose stochastic scattering events pushed them to higher energies. Larger simulation domains are needed to determine what influence the high-energy particle pile-up could have on the particle spectra at lower energies. 

\begin{figure}
    \centering
    \includegraphics[width=0.45\textwidth]{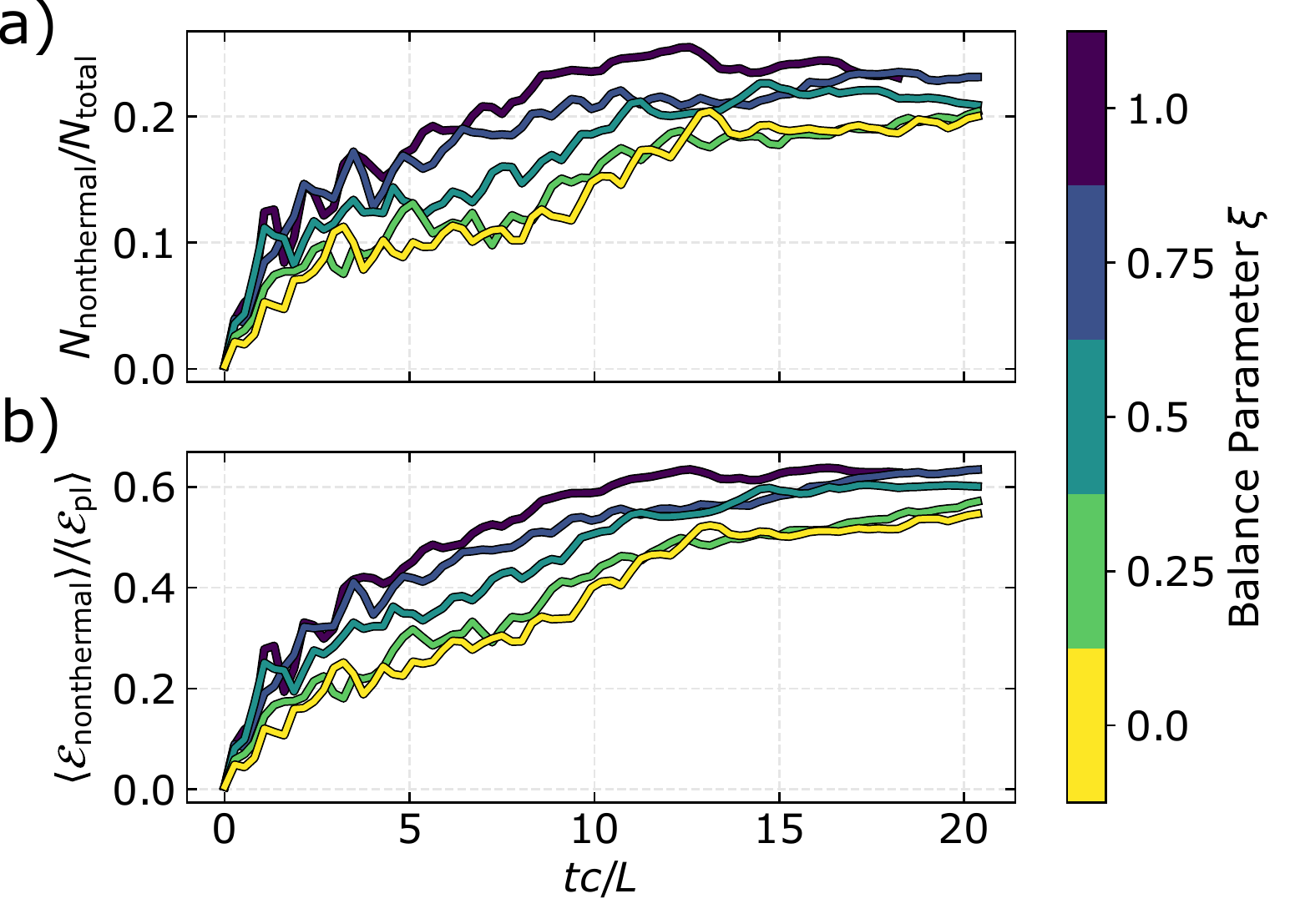}
    \caption{The partition of plasma energy~$\langle\mathcal{E}_{\rm pl}\rangle$ into thermal and nonthermal components shows a moderate increase with the balance parameter at any given time. Both the fraction of particles with nonthermal energies (a) and the fraction of total plasma energy density~$\langle\mathcal{E}_{\rm pl}\rangle$ contained in such particles (b) are calculated by fitting a thermal Maxwell-J\"uttner function to the low- and medium-energy particle distribution at each time and subtracting the fit from the total particle distribution. Colors are the same as in Fig.~\ref{fig:grid}.}
    \label{fig:nonthermalfraction}
\end{figure}

The partition of the plasma particles' energy~$\langle\mathcal{E}_{\rm pl}\rangle(t)$ into thermal and nonthermal components further demonstrates that nonthermal particles are energetically important in the system. The fraction of nonthermal particles is calculated by subtracting the Maxwell-J\"uttner distribution that best fits the total, box-averaged particle distribution up to the peak Lorentz factor from the total distribution function. This fraction reaches 20\% of the total number of particles at~$20~L/c$ for the~$\xi=0$ case (Fig.~\ref{fig:nonthermalfraction}a). The balanced case's fraction of nonthermal particles is larger, reaching 25\% of the total number of particles by the same time. At~$20~L/c$, the energy in these particles comprises 55\% of the total plasma energy in the imbalanced case, as compared to 65\% for the balanced case (Fig.~\ref{fig:nonthermalfraction}b). At equivalent times, the nonthermal fractions of particles and energy do not vary more than 2\% across balance parameters (Table~\ref{tab:teq}), suggesting that particle acceleration by imbalanced turbulence is just as efficient as acceleration by balanced turbulence. For comparison, the fractions for the balanced simulations are slightly smaller than those found for similar simulations of electron-ion plasmas in the relativistically-hot limit~\citep{Zhdankin2019}. 

Though highly idealized, quasi-linear theory can explain many aspects of turbulent NTPA. In particular, the treatment of NTPA as a diffusion in momentum space~\citep{Schlickeiser1989,Chandran2000} has been justified by measurements of the momentum diffusion coefficient in PIC simulations of balanced turbulence~\citep{Comisso2019,Wong2020} and improved by considering resonance broadening~\citep{Demidem2020}. The original models suggest that the diffusion coefficient scales as $1-\tilde H_c^2$, which is supported by test-particle simulations of imbalanced MHD turbulence when parallel acceleration is negligible~\citep{Teaca2014}. Our simulations show that imbalance increases the acceleration timescale for NTPA, which is broadly consistent with a decrease in the diffusion coefficient. It is not clear why this increased acceleration timescale does not affect the power-law index.


\section{Conclusions}\label{sec:conclusions}

In this study we investigate, for the first time, imbalanced kinetic turbulence in a collisionless, magnetized, relativistically hot plasma. Using 3D PIC simulations, we simulate a pair plasma driven by large-scale external currents, creating Alfv\'en waves propagating parallel and anti-parallel to the background magnetic field with different amplitudes. We demonstrate the formation of a turbulent cascade with a similar power-law index for all values of the balance parameter covered by the simulations (Section~\ref{ssec:turbcascade}). We find that the energy injected into the plasma by the external driving is not only converted into internal energy through small-scale dissipative processes (Section~\ref{ssec:energypartition}), but also drives net bulk motion of the plasma (Section~\ref{ssec:net}). This efficient transfer of momentum to the plasma appears as a relativistic effect, resulting in a net plasma velocity~$\Gamma_{\rm net} v_{\rm net}\sim \delta\sigma~v_{A0}$. We also find efficient particle acceleration over two decades of particle Lorentz factor even for our most imbalanced turbulence (Section~\ref{ssec:ntpa}). 

Our results on imbalanced turbulence should apply to high-energy astrophysical systems with asymmetric energy injection, such as accretion disk coronae, relativistic jets, and pulsar wind nebulae. We find that NTPA remains efficient in imbalanced turbulence, meaning that particle acceleration models developed for balanced turbulence are still applicable to astrophysical systems with asymmetric energy injection. In addition, our finding that the momentum from the driven Alfv\'en waves efficiently transfers to the plasma itself constitutes a new mechanism for propelling winds from, for instance, the surface of a turbulent accretion disk. This efficient momentum transfer could also amplify existing astrophysical outflows.

This work represents an important step in studying global properties of imbalanced turbulence in collisionless plasmas. It demonstrates a method for driving imbalanced turbulence in PIC simulations and develops diagnostics to study the unique aspects of imbalanced turbulence, including net motion of the plasma. Our study has revealed a number of factors that could influence the development of imbalanced turbulence and should be further explored: the driving mechanism, the amplitude of magnetic field fluctuations, and the plasma magnetization, to name a few. Our main findings of efficient NTPA and efficient momentum transfer merit further investigation: how are the Fokker-Planck momentum diffusion and advection coefficients for NTPA modified by imbalance? How does the momentum transfer manifest in a more realistic system with density gradients? Thus far, our finding of a net flow is a momentum-transfer mechanism, not a wind-launching mechanism. More work is needed to determine how the transfer efficiency changes with~$\sigma$ and whether the wind comprises the thermal bulk of particles or nonthermal particles. Simulations of imbalanced turbulence in nonrelativistic, semi-relativistic, and trans-relativistic electron-ion plasmas, particularly relevant to accretion flows, will also be important for understanding the fraction of energy that heats electrons. Understanding these aspects of imbalanced turbulence will aid in modeling astrophysical systems with asymmetric energy injection, such as accretion disk coronae, relativistic jets, and pulsar wind nebulae.

\section*{Acknowledgements}

AMH acknowledges support from the National Science Foundation Graduate Research Fellowship Program under Grant No. DGE 1650115. VZ acknowledges support from NASA Hubble Fellowship grant HST-HF2-51426.001-A. This work was also supported by NASA ATP grants NNX17AK57 and 80NSSC20K0545, and NSF Grants AST-1806084 and AST-1903335. 

This work used the Extreme Science and Engineering Discovery Environment (XSEDE), which is supported by National Science Foundation grant number ACI-1548562~\citep{xsede}. We used the XSEDE resources Stampede2 and Ranch at the Texas Advanced Computing Center (TACC) through allocations PHY-140041 and PHY-160032.

\section*{Data Availability}
The simulation data underlying this article were generated at the XSEDE/TACC Stampede2 supercomputer and are archived at the
TACC/Ranch storage facility. As long as the data remain in the
archive, they will be shared on reasonable request to the corresponding author.



\bibliographystyle{mnras}
\bibliography{refs} 

\begin{thebibliography}{}
\makeatletter
\relax
\def\mn@urlcharsother{\let\do\@makeother \do\$\do\&\do\#\do\^\do\_\do\%\do\~}
\def\mn@doi{\begingroup\mn@urlcharsother \@ifnextchar [ {\mn@doi@}
  {\mn@doi@[]}}
\def\mn@doi@[#1]#2{\def\@tempa{#1}\ifx\@tempa\@empty \href
  {http://dx.doi.org/#2} {doi:#2}\else \href {http://dx.doi.org/#2} {#1}\fi
  \endgroup}
\def\mn@eprint#1#2{\mn@eprint@#1:#2::\@nil}
\def\mn@eprint@arXiv#1{\href {http://arxiv.org/abs/#1} {{\tt arXiv:#1}}}
\def\mn@eprint@dblp#1{\href {http://dblp.uni-trier.de/rec/bibtex/#1.xml}
  {dblp:#1}}
\def\mn@eprint@#1:#2:#3:#4\@nil{\def\@tempa {#1}\def\@tempb {#2}\def\@tempc
  {#3}\ifx \@tempc \@empty \let \@tempc \@tempb \let \@tempb \@tempa \fi \ifx
  \@tempb \@empty \def\@tempb {arXiv}\fi \@ifundefined
  {mn@eprint@\@tempb}{\@tempb:\@tempc}{\expandafter \expandafter \csname
  mn@eprint@\@tempb\endcsname \expandafter{\@tempc}}}

\bibitem[\protect\citeauthoryear{{Abdo} et~al.,}{{Abdo}
  et~al.}{2009}]{Abdo2009}
{Abdo} A.~A.,  et~al., 2009, \mn@doi [\prl] {10.1103/PhysRevLett.102.181101},
  \href {https://ui.adsabs.harvard.edu/abs/2009PhRvL.102r1101A} {102, 181101}

\bibitem[\protect\citeauthoryear{{Aleksi{\'c}} et~al.,}{{Aleksi{\'c}}
  et~al.}{2015}]{Aleksic2015}
{Aleksi{\'c}} J.,  et~al., 2015, \mn@doi [Journal of High Energy Astrophysics]
  {10.1016/j.jheap.2015.01.002}, \href
  {https://ui.adsabs.harvard.edu/abs/2015JHEAp...5...30A} {5, 30}

\bibitem[\protect\citeauthoryear{{Alves}, {Zrake}  \& {Fiuza}}{{Alves}
  et~al.}{2018}]{Alves2018}
{Alves} E.~P.,  {Zrake} J.,   {Fiuza} F.,  2018, \mn@doi [\prl]
  {10.1103/PhysRevLett.121.245101}, \href
  {https://ui.adsabs.harvard.edu/abs/2018PhRvL.121x5101A} {121, 245101}

\bibitem[\protect\citeauthoryear{{Balbus} \& {Hawley}}{{Balbus} \&
  {Hawley}}{1998}]{Balbus1998}
{Balbus} S.~A.,  {Hawley} J.~F.,  1998, \mn@doi [Reviews of Modern Physics]
  {10.1103/RevModPhys.70.1}, \href
  {https://ui.adsabs.harvard.edu/abs/1998RvMP...70....1B} {70, 1}

\bibitem[\protect\citeauthoryear{{Beresnyak} \& {Lazarian}}{{Beresnyak} \&
  {Lazarian}}{2008}]{Beresnyak2008}
{Beresnyak} A.,  {Lazarian} A.,  2008, \mn@doi [\apj] {10.1086/589428}, \href
  {https://ui.adsabs.harvard.edu/abs/2008ApJ...682.1070B} {682, 1070}

\bibitem[\protect\citeauthoryear{{Beresnyak} \& {Lazarian}}{{Beresnyak} \&
  {Lazarian}}{2009}]{Beresnyak2009}
{Beresnyak} A.,  {Lazarian} A.,  2009, \mn@doi [\apj]
  {10.1088/0004-637X/702/1/460}, \href
  {https://ui.adsabs.harvard.edu/abs/2009ApJ...702..460B} {702, 460}

\bibitem[\protect\citeauthoryear{{Beresnyak} \& {Lazarian}}{{Beresnyak} \&
  {Lazarian}}{2010}]{Beresnyak2010}
{Beresnyak} A.,  {Lazarian} A.,  2010, \mn@doi [\apjl]
  {10.1088/2041-8205/722/1/L110}, \href
  {https://ui.adsabs.harvard.edu/abs/2010ApJ...722L.110B} {722, L110}

\bibitem[\protect\citeauthoryear{{Boldyrev}}{{Boldyrev}}{2006}]{Boldyrev2006}
{Boldyrev} S.,  2006, \mn@doi [\prl] {10.1103/PhysRevLett.96.115002}, \href
  {https://ui.adsabs.harvard.edu/abs/2006PhRvL..96k5002B} {96, 115002}

\bibitem[\protect\citeauthoryear{{Boldyrev} \& {Perez}}{{Boldyrev} \&
  {Perez}}{2009}]{Boldyrev2009}
{Boldyrev} S.,  {Perez} J.~C.,  2009, \mn@doi [\prl]
  {10.1103/PhysRevLett.103.225001}, \href
  {https://ui.adsabs.harvard.edu/abs/2009PhRvL.103v5001B} {103, 225001}

\bibitem[\protect\citeauthoryear{{Boldyrev}, {Perez}, {Borovsky}  \&
  {Podesta}}{{Boldyrev} et~al.}{2011}]{Boldyrev2011}
{Boldyrev} S.,  {Perez} J.~C.,  {Borovsky} J.~E.,   {Podesta} J.~J.,  2011,
  \mn@doi [\apjl] {10.1088/2041-8205/741/1/L19}, \href
  {https://ui.adsabs.harvard.edu/abs/2011ApJ...741L..19B} {741, L19}

\bibitem[\protect\citeauthoryear{{Cerutti}, {Werner}, {Uzdensky}  \&
  {Begelman}}{{Cerutti} et~al.}{2013}]{Cerutti2013}
{Cerutti} B.,  {Werner} G.~R.,  {Uzdensky} D.~A.,   {Begelman} M.~C.,  2013,
  \mn@doi [\apj] {10.1088/0004-637X/770/2/147}, \href
  {https://ui.adsabs.harvard.edu/abs/2013ApJ...770..147C} {770, 147}

\bibitem[\protect\citeauthoryear{{Chandran}}{{Chandran}}{2000}]{Chandran2000}
{Chandran} B. D.~G.,  2000, \mn@doi [\prl] {10.1103/PhysRevLett.85.4656}, \href
  {https://ui.adsabs.harvard.edu/abs/2000PhRvL..85.4656C} {85, 4656}

\bibitem[\protect\citeauthoryear{{Chandran}}{{Chandran}}{2008}]{Chandran2008}
{Chandran} B. D.~G.,  2008, \mn@doi [\apj] {10.1086/589432}, \href
  {https://ui.adsabs.harvard.edu/abs/2008ApJ...685..646C} {685, 646}

\bibitem[\protect\citeauthoryear{{Chandran}, {Foucart}  \&
  {Tchekhovskoy}}{{Chandran} et~al.}{2018}]{Chandran2018}
{Chandran} B. D.~G.,  {Foucart} F.,   {Tchekhovskoy} A.,  2018, \mn@doi
  [Journal of Plasma Physics] {10.1017/S0022377818000387}, \href
  {https://ui.adsabs.harvard.edu/abs/2018JPlPh..84c9010C} {84, 905840310}

\bibitem[\protect\citeauthoryear{{Chen}}{{Chen}}{2016}]{Chen2016}
{Chen} C.~H.~K.,  2016, \mn@doi [Journal of Plasma Physics]
  {10.1017/S0022377816001124}, \href
  {https://ui.adsabs.harvard.edu/abs/2016JPlPh..82f5302C} {82, 535820602}

\bibitem[\protect\citeauthoryear{{Chen}, {Bale}, {Salem}  \& {Maruca}}{{Chen}
  et~al.}{2013}]{Chen2013a}
{Chen} C.~H.~K.,  {Bale} S.~D.,  {Salem} C.~S.,   {Maruca} B.~A.,  2013,
  \mn@doi [\apj] {10.1088/0004-637X/770/2/125}, \href
  {https://ui.adsabs.harvard.edu/abs/2013ApJ...770..125C} {770, 125}

\bibitem[\protect\citeauthoryear{{Cho} \& {Kim}}{{Cho} \&
  {Kim}}{2016}]{Cho2016}
{Cho} J.,  {Kim} H.,  2016, \mn@doi [Journal of Geophysical Research (Space
  Physics)] {10.1002/2016JA022364}, \href
  {https://ui.adsabs.harvard.edu/abs/2016JGRA..121.6157C} {121, 6157}

\bibitem[\protect\citeauthoryear{{Cho} \& {Lazarian}}{{Cho} \&
  {Lazarian}}{2014}]{Cho2014}
{Cho} J.,  {Lazarian} A.,  2014, \mn@doi [\apj] {10.1088/0004-637X/780/1/30},
  \href {https://ui.adsabs.harvard.edu/abs/2014ApJ...780...30C} {780, 30}

\bibitem[\protect\citeauthoryear{{Comisso} \& {Sironi}}{{Comisso} \&
  {Sironi}}{2018}]{Comisso2018}
{Comisso} L.,  {Sironi} L.,  2018, \mn@doi [\prl]
  {10.1103/PhysRevLett.121.255101}, \href
  {https://ui.adsabs.harvard.edu/abs/2018PhRvL.121y5101C} {121, 255101}

\bibitem[\protect\citeauthoryear{{Comisso} \& {Sironi}}{{Comisso} \&
  {Sironi}}{2019}]{Comisso2019}
{Comisso} L.,  {Sironi} L.,  2019, \mn@doi [\apj] {10.3847/1538-4357/ab4c33},
  \href {https://ui.adsabs.harvard.edu/abs/2019ApJ...886..122C} {886, 122}

\bibitem[\protect\citeauthoryear{{Davelaar}, {Philippov}, {Bromberg}  \&
  {Singh}}{{Davelaar} et~al.}{2020}]{Davelaar2020}
{Davelaar} J.,  {Philippov} A.~A.,  {Bromberg} O.,   {Singh} C.~B.,  2020,
  \mn@doi [\apjl] {10.3847/2041-8213/ab95a2}, \href
  {https://ui.adsabs.harvard.edu/abs/2020ApJ...896L..31D} {896, L31}

\bibitem[\protect\citeauthoryear{{Demidem}, {Lemoine}  \& {Casse}}{{Demidem}
  et~al.}{2020}]{Demidem2020}
{Demidem} C.,  {Lemoine} M.,   {Casse} F.,  2020, \mn@doi [\prd]
  {10.1103/PhysRevD.102.023003}, \href
  {https://ui.adsabs.harvard.edu/abs/2020PhRvD.102b3003D} {102, 023003}

\bibitem[\protect\citeauthoryear{Elsasser}{Elsasser}{1950}]{Elsasser1950}
Elsasser W.~M.,  1950, \mn@doi [Phys. Rev.] {10.1103/PhysRev.79.183}, 79, 183

\bibitem[\protect\citeauthoryear{{Galtier}, {Nazarenko}, {Newell}  \&
  {Pouquet}}{{Galtier} et~al.}{2000}]{Galtier2000}
{Galtier} S.,  {Nazarenko} S.~V.,  {Newell} A.~C.,   {Pouquet} A.,  2000,
  \mn@doi [Journal of Plasma Physics] {10.1017/S0022377899008284}, \href
  {https://ui.adsabs.harvard.edu/abs/2000JPlPh..63..447G} {63, 447}

\bibitem[\protect\citeauthoryear{{Gogoberidze} \& {Voitenko}}{{Gogoberidze} \&
  {Voitenko}}{2020}]{Gogoberidze2020}
{Gogoberidze} G.,  {Voitenko} Y.~M.,  2020, \mn@doi [\apss]
  {10.1007/s10509-020-03865-8}, \href
  {https://ui.adsabs.harvard.edu/abs/2020Ap&SS.365..149G} {365, 149}

\bibitem[\protect\citeauthoryear{{Goldreich} \& {Sridhar}}{{Goldreich} \&
  {Sridhar}}{1995}]{GS1995}
{Goldreich} P.,  {Sridhar} S.,  1995, \mn@doi [\apj] {10.1086/175121}, \href
  {https://ui.adsabs.harvard.edu/abs/1995ApJ...438..763G} {438, 763}

\bibitem[\protect\citeauthoryear{{Gro{\v{s}}elj}, {Mallet}, {Loureiro}  \&
  {Jenko}}{{Gro{\v{s}}elj} et~al.}{2018}]{Groselj2018}
{Gro{\v{s}}elj} D.,  {Mallet} A.,  {Loureiro} N.~F.,   {Jenko} F.,  2018,
  \mn@doi [\prl] {10.1103/PhysRevLett.120.105101}, \href
  {https://ui.adsabs.harvard.edu/abs/2018PhRvL.120j5101G} {120, 105101}

\bibitem[\protect\citeauthoryear{{Hester}}{{Hester}}{2008}]{Hester2008}
{Hester} J.~J.,  2008, \mn@doi [\araa]
  {10.1146/annurev.astro.45.051806.110608}, \href
  {https://ui.adsabs.harvard.edu/abs/2008ARA&A..46..127H} {46, 127}

\bibitem[\protect\citeauthoryear{{Hoshino}}{{Hoshino}}{2013}]{Hoshino2013}
{Hoshino} M.,  2013, \mn@doi [\apj] {10.1088/0004-637X/773/2/118}, \href
  {https://ui.adsabs.harvard.edu/abs/2013ApJ...773..118H} {773, 118}

\bibitem[\protect\citeauthoryear{{Iroshnikov}}{{Iroshnikov}}{1964}]{Iroshnikov1964}
{Iroshnikov} P.~S.,  1964, \sovast, \href
  {https://ui.adsabs.harvard.edu/abs/1964SvA.....7..566I} {7, 566}

\bibitem[\protect\citeauthoryear{{Kim} \& {Cho}}{{Kim} \&
  {Cho}}{2015}]{Kim2015}
{Kim} H.,  {Cho} J.,  2015, \mn@doi [\apj] {10.1088/0004-637X/801/2/75}, \href
  {https://ui.adsabs.harvard.edu/abs/2015ApJ...801...75K} {801, 75}

\bibitem[\protect\citeauthoryear{{Kraichnan}}{{Kraichnan}}{1965}]{Kraichnan1965}
{Kraichnan} R.~H.,  1965, \mn@doi [Physics of Fluids] {10.1063/1.1761412},
  \href {https://ui.adsabs.harvard.edu/abs/1965PhFl....8.1385K} {8, 1385}

\bibitem[\protect\citeauthoryear{{Kunz}, {Stone}  \& {Quataert}}{{Kunz}
  et~al.}{2016}]{Kunz2016}
{Kunz} M.~W.,  {Stone} J.~M.,   {Quataert} E.,  2016, \mn@doi [\prl]
  {10.1103/PhysRevLett.117.235101}, \href
  {https://ui.adsabs.harvard.edu/abs/2016PhRvL.117w5101K} {117, 235101}

\bibitem[\protect\citeauthoryear{{Lithwick} \& {Goldreich}}{{Lithwick} \&
  {Goldreich}}{2001}]{Lithwick2001}
{Lithwick} Y.,  {Goldreich} P.,  2001, \mn@doi [\apj] {10.1086/323470}, \href
  {https://ui.adsabs.harvard.edu/abs/2001ApJ...562..279L} {562, 279}

\bibitem[\protect\citeauthoryear{{Lithwick}, {Goldreich}  \&
  {Sridhar}}{{Lithwick} et~al.}{2007}]{Lithwick2007}
{Lithwick} Y.,  {Goldreich} P.,   {Sridhar} S.,  2007, \mn@doi [\apj]
  {10.1086/509884}, \href
  {https://ui.adsabs.harvard.edu/abs/2007ApJ...655..269L} {655, 269}

\bibitem[\protect\citeauthoryear{{Mason}, {Perez}, {Boldyrev}  \&
  {Cattaneo}}{{Mason} et~al.}{2012}]{Mason2012}
{Mason} J.,  {Perez} J.~C.,  {Boldyrev} S.,   {Cattaneo} F.,  2012, \mn@doi
  [Physics of Plasmas] {10.1063/1.3694123}, \href
  {https://ui.adsabs.harvard.edu/abs/2012PhPl...19e5902M} {19, 055902}

\bibitem[\protect\citeauthoryear{{Meyrand}, {Squire}, {Schekochihin}  \&
  {Dorland}}{{Meyrand} et~al.}{2021}]{Meyrand2020}
{Meyrand} R.,  {Squire} J.,  {Schekochihin} A.~A.,   {Dorland} W.,  2021,
  \mn@doi [Journal of Plasma Physics] {10.1017/S0022377821000489}, \href
  {https://ui.adsabs.harvard.edu/abs/2021JPlPh..87c5301M} {87, 535870301}

\bibitem[\protect\citeauthoryear{{Miloshevich}, {Passot}  \&
  {Sulem}}{{Miloshevich} et~al.}{2020}]{Miloshevich2020}
{Miloshevich} G.,  {Passot} T.,   {Sulem} P.~L.,  2020, \mn@doi [\apjl]
  {10.3847/2041-8213/ab60b1}, \href
  {https://ui.adsabs.harvard.edu/abs/2020ApJ...888L...7M} {888, L7}

\bibitem[\protect\citeauthoryear{{Passot} \& {Sulem}}{{Passot} \&
  {Sulem}}{2019}]{Passot2019}
{Passot} T.,  {Sulem} P.~L.,  2019, \mn@doi [Journal of Plasma Physics]
  {10.1017/S0022377819000187}, \href
  {https://ui.adsabs.harvard.edu/abs/2019JPlPh..85c9001P} {85, 905850301}

\bibitem[\protect\citeauthoryear{{Perez} \& {Boldyrev}}{{Perez} \&
  {Boldyrev}}{2009}]{Perez2009}
{Perez} J.~C.,  {Boldyrev} S.,  2009, \mn@doi [\prl]
  {10.1103/PhysRevLett.102.025003}, \href
  {https://ui.adsabs.harvard.edu/abs/2009PhRvL.102b5003P} {102, 025003}

\bibitem[\protect\citeauthoryear{{Perez} \& {Boldyrev}}{{Perez} \&
  {Boldyrev}}{2010a}]{PB2010b}
{Perez} J.~C.,  {Boldyrev} S.,  2010a, \mn@doi [Physics of Plasmas]
  {10.1063/1.3396370}, \href
  {https://ui.adsabs.harvard.edu/abs/2010PhPl...17e5903P} {17, 055903}

\bibitem[\protect\citeauthoryear{{Perez} \& {Boldyrev}}{{Perez} \&
  {Boldyrev}}{2010b}]{PB2010a}
{Perez} J.~C.,  {Boldyrev} S.,  2010b, \mn@doi [\apjl]
  {10.1088/2041-8205/710/1/L63}, \href
  {https://ui.adsabs.harvard.edu/abs/2010ApJ...710L..63P} {710, L63}

\bibitem[\protect\citeauthoryear{{Perez}, {Mason}, {Boldyrev}  \&
  {Cattaneo}}{{Perez} et~al.}{2012}]{Perez2012}
{Perez} J.~C.,  {Mason} J.,  {Boldyrev} S.,   {Cattaneo} F.,  2012, \mn@doi
  [Physical Review X] {10.1103/PhysRevX.2.041005}, \href
  {https://ui.adsabs.harvard.edu/abs/2012PhRvX...2d1005P} {2, 041005}

\bibitem[\protect\citeauthoryear{{Remillard} \& {McClintock}}{{Remillard} \&
  {McClintock}}{2006}]{Remillard2006}
{Remillard} R.~A.,  {McClintock} J.~E.,  2006, \mn@doi [\araa]
  {10.1146/annurev.astro.44.051905.092532}, \href
  {https://ui.adsabs.harvard.edu/abs/2006ARA&A..44...49R} {44, 49}

\bibitem[\protect\citeauthoryear{{Ripperda} et~al.,}{{Ripperda}
  et~al.}{2021}]{Ripperda2021}
{Ripperda} B.,  et~al., 2021, arXiv e-prints, \href
  {https://ui.adsabs.harvard.edu/abs/2021arXiv210501145R} {p. arXiv:2105.01145}

\bibitem[\protect\citeauthoryear{{Riquelme}, {Quataert}, {Sharma}  \&
  {Spitkovsky}}{{Riquelme} et~al.}{2012}]{Riquelme2012}
{Riquelme} M.~A.,  {Quataert} E.,  {Sharma} P.,   {Spitkovsky} A.,  2012,
  \mn@doi [\apj] {10.1088/0004-637X/755/1/50}, \href
  {https://ui.adsabs.harvard.edu/abs/2012ApJ...755...50R} {755, 50}

\bibitem[\protect\citeauthoryear{{Schekochihin}}{{Schekochihin}}{2020}]{Schekochihin2020}
{Schekochihin} A.~A.,  2020, arXiv e-prints, \href
  {https://ui.adsabs.harvard.edu/abs/2020arXiv201000699S} {p. arXiv:2010.00699}

\bibitem[\protect\citeauthoryear{{Schekochihin}, {Cowley}, {Dorland},
  {Hammett}, {Howes}, {Quataert}  \& {Tatsuno}}{{Schekochihin}
  et~al.}{2009}]{Schekochihin2009}
{Schekochihin} A.~A.,  {Cowley} S.~C.,  {Dorland} W.,  {Hammett} G.~W.,
  {Howes} G.~G.,  {Quataert} E.,   {Tatsuno} T.,  2009, \mn@doi [\apjs]
  {10.1088/0067-0049/182/1/310}, \href
  {https://ui.adsabs.harvard.edu/abs/2009ApJS..182..310S} {182, 310}

\bibitem[\protect\citeauthoryear{{Schlickeiser}}{{Schlickeiser}}{1989}]{Schlickeiser1989}
{Schlickeiser} R.,  1989, \mn@doi [\apj] {10.1086/167009}, \href
  {https://ui.adsabs.harvard.edu/abs/1989ApJ...336..243S} {336, 243}

\bibitem[\protect\citeauthoryear{{Shakura} \& {Sunyaev}}{{Shakura} \&
  {Sunyaev}}{1973}]{SS1973}
{Shakura} N.~I.,  {Sunyaev} R.~A.,  1973, \aap, \href
  {https://ui.adsabs.harvard.edu/abs/1973A&A....24..337S} {500, 33}

\bibitem[\protect\citeauthoryear{{Sparks}, {Biretta}  \& {Macchetto}}{{Sparks}
  et~al.}{1996}]{Sparks1996}
{Sparks} W.~B.,  {Biretta} J.~A.,   {Macchetto} F.,  1996, \mn@doi [\apj]
  {10.1086/178141}, \href
  {https://ui.adsabs.harvard.edu/abs/1996ApJ...473..254S} {473, 254}

\bibitem[\protect\citeauthoryear{{Takamoto} \& {Lazarian}}{{Takamoto} \&
  {Lazarian}}{2016}]{Takamoto2016}
{Takamoto} M.,  {Lazarian} A.,  2016, \mn@doi [\apjl]
  {10.3847/2041-8205/831/2/L11}, \href
  {https://ui.adsabs.harvard.edu/abs/2016ApJ...831L..11T} {831, L11}

\bibitem[\protect\citeauthoryear{{Teaca}, {Weidl}, {Jenko}  \&
  {Schlickeiser}}{{Teaca} et~al.}{2014}]{Teaca2014}
{Teaca} B.,  {Weidl} M.~S.,  {Jenko} F.,   {Schlickeiser} R.,  2014, \mn@doi
  [\pre] {10.1103/PhysRevE.90.021101}, \href
  {https://ui.adsabs.harvard.edu/abs/2014PhRvE..90b1101T} {90, 021101}

\bibitem[\protect\citeauthoryear{{TenBarge}, {Howes}, {Dorland}  \&
  {Hammett}}{{TenBarge} et~al.}{2014}]{TenBarge2014}
{TenBarge} J.~M.,  {Howes} G.~G.,  {Dorland} W.,   {Hammett} G.~W.,  2014,
  \mn@doi [Computer Physics Communications] {10.1016/j.cpc.2013.10.022}, \href
  {https://ui.adsabs.harvard.edu/abs/2014CoPhC.185..578T} {185, 578}

\bibitem[\protect\citeauthoryear{{TenBarge} et~al.,}{{TenBarge}
  et~al.}{2021}]{TenBarge2021}
{TenBarge} J.~M.,  et~al., 2021, arXiv e-prints, \href
  {https://ui.adsabs.harvard.edu/abs/2021arXiv210501146T} {p. arXiv:2105.01146}

\bibitem[\protect\citeauthoryear{{Thompson} \& {Blaes}}{{Thompson} \&
  {Blaes}}{1998}]{Thompson1998}
{Thompson} C.,  {Blaes} O.,  1998, \mn@doi [\prd] {10.1103/PhysRevD.57.3219},
  \href {https://ui.adsabs.harvard.edu/abs/1998PhRvD..57.3219T} {57, 3219}

\bibitem[\protect\citeauthoryear{Towns et~al.,}{Towns et~al.}{2014}]{xsede}
Towns J.,  et~al., 2014, \mn@doi [Computing in Science \& Engineering]
  {10.1109/MCSE.2014.80}, 16, 62

\bibitem[\protect\citeauthoryear{{Voitenko} \& {De Keyser}}{{Voitenko} \& {De
  Keyser}}{2016}]{Voitenko2016}
{Voitenko} Y.,  {De Keyser} J.,  2016, \mn@doi [\apjl]
  {10.3847/2041-8205/832/2/L20}, \href
  {https://ui.adsabs.harvard.edu/abs/2016ApJ...832L..20V} {832, L20}

\bibitem[\protect\citeauthoryear{{Wang}, {Boldyrev}  \& {Perez}}{{Wang}
  et~al.}{2011}]{Wang2011}
{Wang} Y.,  {Boldyrev} S.,   {Perez} J.~C.,  2011, \mn@doi [\apjl]
  {10.1088/2041-8205/740/2/L36}, \href
  {https://ui.adsabs.harvard.edu/abs/2011ApJ...740L..36W} {740, L36}

\bibitem[\protect\citeauthoryear{{Wong}, {Zhdankin}, {Uzdensky}, {Werner}  \&
  {Begelman}}{{Wong} et~al.}{2020}]{Wong2020}
{Wong} K.,  {Zhdankin} V.,  {Uzdensky} D.~A.,  {Werner} G.~R.,   {Begelman}
  M.~C.,  2020, \mn@doi [\apjl] {10.3847/2041-8213/ab8122}, \href
  {https://ui.adsabs.harvard.edu/abs/2020ApJ...893L...7W} {893, L7}

\bibitem[\protect\citeauthoryear{{Zhdankin}}{{Zhdankin}}{2021}]{Zhdankin2021}
{Zhdankin} V.,  2021, arXiv e-prints, \href
  {https://ui.adsabs.harvard.edu/abs/2021arXiv210600743Z} {p. arXiv:2106.00743}

\bibitem[\protect\citeauthoryear{{Zhdankin}, {Werner}, {Uzdensky}  \&
  {Begelman}}{{Zhdankin} et~al.}{2017}]{Zhdankin2017}
{Zhdankin} V.,  {Werner} G.~R.,  {Uzdensky} D.~A.,   {Begelman} M.~C.,  2017,
  \mn@doi [\prl] {10.1103/PhysRevLett.118.055103}, \href
  {https://ui.adsabs.harvard.edu/abs/2017PhRvL.118e5103Z} {118, 055103}

\bibitem[\protect\citeauthoryear{{Zhdankin}, {Uzdensky}, {Werner}  \&
  {Begelman}}{{Zhdankin} et~al.}{2018a}]{Zhdankin2018}
{Zhdankin} V.,  {Uzdensky} D.~A.,  {Werner} G.~R.,   {Begelman} M.~C.,  2018a,
  \mn@doi [\mnras] {10.1093/mnras/stx2883}, \href
  {https://ui.adsabs.harvard.edu/abs/2018MNRAS.474.2514Z} {474, 2514}

\bibitem[\protect\citeauthoryear{{Zhdankin}, {Uzdensky}, {Werner}  \&
  {Begelman}}{{Zhdankin} et~al.}{2018b}]{Zhdankin2018b}
{Zhdankin} V.,  {Uzdensky} D.~A.,  {Werner} G.~R.,   {Begelman} M.~C.,  2018b,
  \mn@doi [\apjl] {10.3847/2041-8213/aae88c}, \href
  {https://ui.adsabs.harvard.edu/abs/2018ApJ...867L..18Z} {867, L18}

\bibitem[\protect\citeauthoryear{{Zhdankin}, {Uzdensky}, {Werner}  \&
  {Begelman}}{{Zhdankin} et~al.}{2019}]{Zhdankin2019}
{Zhdankin} V.,  {Uzdensky} D.~A.,  {Werner} G.~R.,   {Begelman} M.~C.,  2019,
  \mn@doi [\prl] {10.1103/PhysRevLett.122.055101}, \href
  {https://ui.adsabs.harvard.edu/abs/2019PhRvL.122e5101Z} {122, 055101}

\makeatother
\end{thebibliography}

\appendix
\section{Dependence on Domain Size} \label{app:boxsize}%
\begin{figure}
    \centering
    \includegraphics[width=0.45\textwidth]{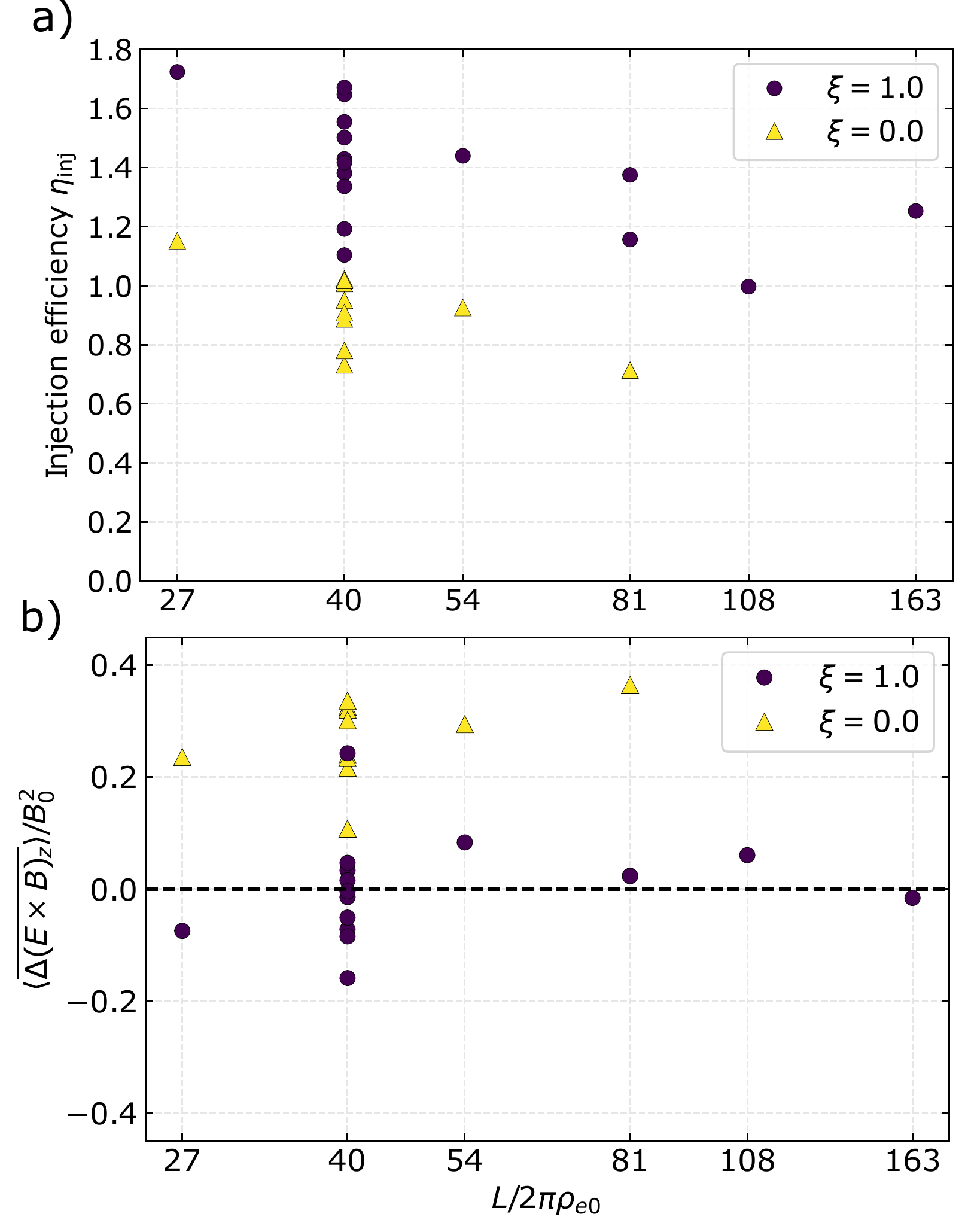}
    \caption{The injection efficiency and Poynting flux along the background magnetic field depend weakly on simulation domain size within statistical variation. When averaged from~$t=5-14~L/c$, the injection efficiency (a) and Poynting flux along the background magnetic field (b) are shown as a function of~$L/2\pi\rho_{e0}$. Both the balanced ($\xi=1.0$; purple circles) and most imbalanced values ($\xi=0$; yellow triangles) are mostly within statistical variation of the~$L/2\pi\rho_{e0}=40.7$ sample of 8 random seeds. The black dashed line indicates zero.}
    \label{fig:boxsize-int-net}
\end{figure}%
\begin{figure}
    \centering
    \includegraphics[width=0.45\textwidth]{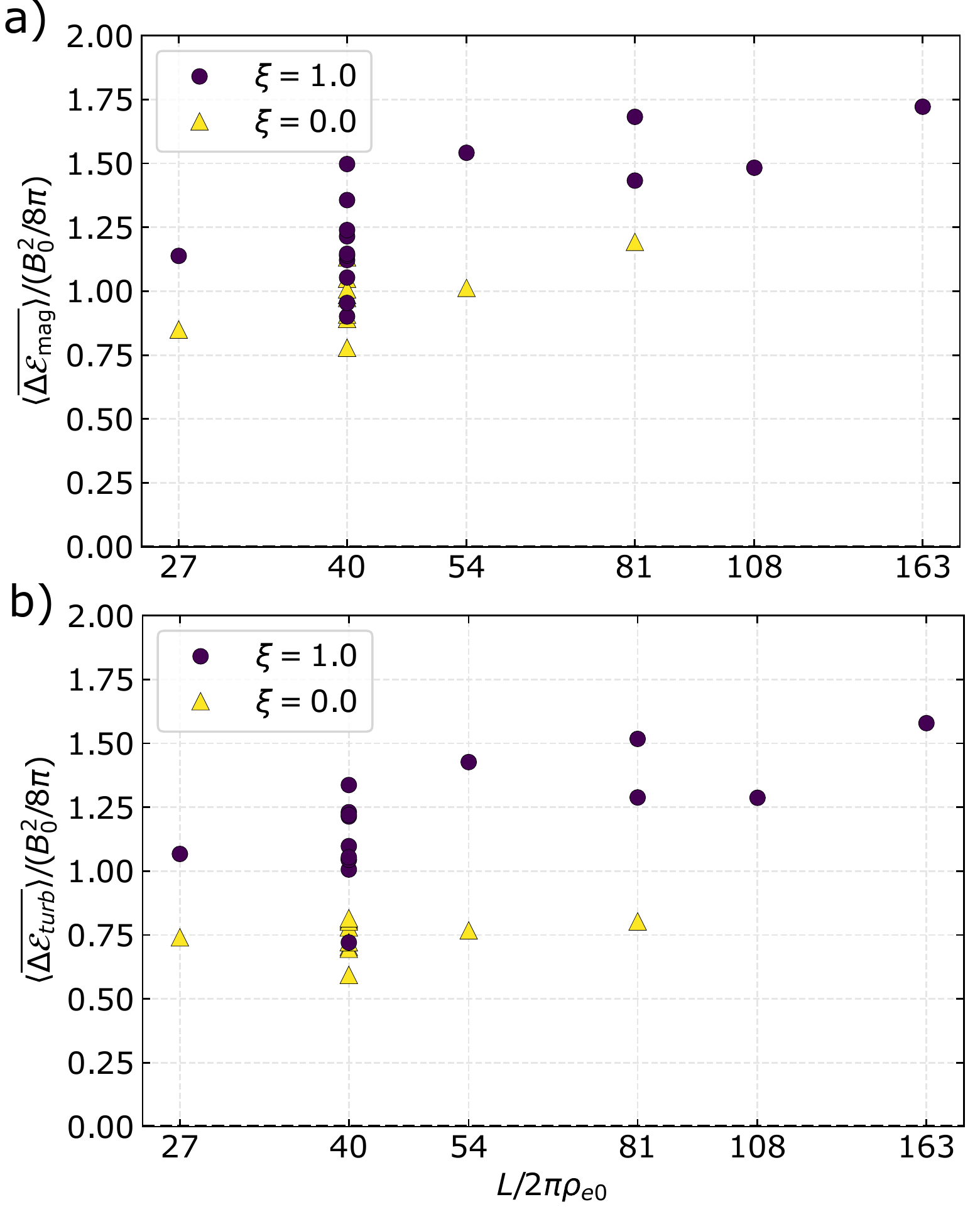}
    \caption{The magnetic and turbulent kinetic densities are weakly dependent on simulation domain size. When averaged from~$t=5-14~L/c$, the magnetic energy density (a) and turbulent kinetic energy density (b) are shown as a function of~$L/2\pi\rho_{e0}$. Both the balanced ($\xi=1.0$; purple circles) and most imbalanced ($\xi=0$; yellow triangles) values are mostly within statistical variation of the~$L/2\pi\rho_{e0}=40.7$ sample of 8 random seeds.}
    \label{fig:boxsize-em-turb}
\end{figure}%
If the ratio~$L/2\pi\rho_{e0}$ is small, the small separation between the characteristic scale of kinetic effects and the system size scale could influence the results presented in Section~\ref{sec:results}. In this appendix, we examine the box-size dependence of representative quantities for extremal values of the balance parameter~$\xi=0.0$ and~$\xi=1.0$ and for~$L/2\pi\rho_{e0}\in\{27.1, 40.7, 54.3, 81.5\}$, corresponding to~$N\in\{256, 384, 512, 768\}$. These results also include three very large simulations of balanced turbulence from~\citet{Zhdankin2018b} with~$L/2\pi\rho_{e0}\in\{81.5, 108.7, 164\}$ ($N=768$,~$1024$, and~$1536$) that are otherwise identical to the other simulations of balanced turbulence. The time-averaging window has been changed from~$10<tc/L<20$ to~$5<tc/L<14$, because~$14~L/c$ is the latest time included in all simulations. Here we focus on the convergence of energetic quantities; for convergence of the balanced turbulence's particle energy spectra with system size, see~\citet{Zhdankin2018b}.

We find that the injection efficiency~$\eta_{\rm inj}$ depends weakly on simulation domain size for both balanced and imbalanced turbulence (Fig.~\ref{fig:boxsize-int-net}a). The simulations with~$\xi=1.0$ and~$L/2\pi\rho_{e0}=81.5$ and 164 domains have slightly lower injection efficiencies than those for the smallest ($L/2\pi\rho_{e0}=40.7$) domains. Kinetic damping of large-scale fluctuations, which would drain energy faster than turbulence alone, may explain the larger~$\eta_{\rm inj}$ for smaller domain sizes. For~$\xi=0$, the~$L/2\pi\rho_{e0}=54.3$ simulation's~$\eta_{\rm inj}$ is within the statistical spread of the~$L/2\pi\rho_{e0}=40.7$ simulations' injection efficiencies, whereas the~$L/2\pi\rho_{e0}=81.5$ injection efficiency is slightly below. The time-averaged Poynting flux shows a weak positive trend with increasing domain size (Fig.~\ref{fig:boxsize-int-net}b).
Though the~$L/2\pi\rho_{e0}=81.5$ domain size for the imbalanced~$\xi=0$ case has a value higher than the largest~$L/2\pi\rho_{e0}=40.7$ value, the difference is only about~$0.05~(cB_0^2/4\pi)$ (about 15\%), within two standard deviations of the statistical variation shown by the~$L/2\pi\rho_{e0}=40.7$ study. 

The turbulent and magnetic energy densities show a slight dependence on simulation domain size (Fig.~\ref{fig:boxsize-em-turb}). Both quantities' values for~$L/2\pi\rho_{e0}\gtrsim54.3$ are consistently about 15\% greater than the largest value of the statistical ensemble of~$L/2\pi\rho_{e0}=40.7$,~$\xi=1.0$ simulations. Though the imbalanced turbulence simulations do not appear to exhibit a trend in turbulent kinetic energy with box size (Fig.~\ref{fig:boxsize-em-turb}b), the~$L/2\pi\rho_{e0}=81.5$,~$\xi=0$ case's value for magnetic energy is noticeably higher (20\%) than the~$L/2\pi\rho_{e0}=54.3$ value, which lies within the statistical spread of the~$L/2\pi\rho_{e0}=40.7$ ensemble study (Fig.~\ref{fig:boxsize-em-turb}a).

\bsp	
\label{lastpage}
\end{document}